\documentclass[aps,reprint,amsmath,amssymb,nofootinbib]{revtex4-1}
\usepackage{amsmath}
\usepackage{amssymb}
\usepackage{graphicx}
\usepackage{natbib}
\renewcommand{\vec}{\mathbf}
\newcommand{\ubLam}{\underline{\bf\Lambda}}
\newcommand{\ubH}{\underline{\bf H}}
\newcommand{\ubQ}{\underline{\bf Q}}
\newcommand{\ubPi}{\underline{\bf \Pi}}
\newcommand{\uPi}{\underline{\Pi}}
\newcommand{\ubGa}{\underline{\bf \Gamma}}
\def\ppd{\perp}

\newcommand{\BVF}{Brunt-V\"ais\"al\"a frequency}
\newcommand{\eq}[1]{(\ref{#1})}
\newcommand{\calL}{\mbox{${\cal L}$}}
\newcommand{\vv}{\vec{v}}
\newcommand{\vn}{\vec{n}}
\newcommand{\vzero}{\vec{0}}
\def\bbbr{{\rm I\!R}}

\def\bbbc{{\mathchoice {\setbox0=\hbox{$\displaystyle\rm C$}\hbox{\hbox
to0pt{\kern0.4\wd0\vrule height0.9\ht0\hss}\box0}}
{\setbox0=\hbox{$\textstyle\rm C$}\hbox{\hbox
to0pt{\kern0.4\wd0\vrule height0.9\ht0\hss}\box0}}
{\setbox0=\hbox{$\scriptstyle\rm C$}\hbox{\hbox
to0pt{\kern0.4\wd0\vrule height0.9\ht0\hss}\box0}}
{\setbox0=\hbox{$\scriptscriptstyle\rm C$}\hbox{\hbox
to0pt{\kern0.4\wd0\vrule height0.9\ht0\hss}\box0}}}}

\newcommand{\beq}{\begin{equation}}
\newcommand{\eeq}{\end{equation}}
\newcommand{\eeqn}[1]{\label{#1}\end{equation}}
\newcommand{\beqan}{\begin{eqnarray}}
\newcommand{\eeqan}[1]{\label{#1}\end{eqnarray}}
\newcommand{\intvol}{ \int_{(V)}\! }
\newcommand{\intsur}{ \int_{(S)}\! }
\newcommand{\dS}{d\vec{S}}
\newcommand{\na}{ \vec{\nabla} }
\newcommand{\andet}{ \qquad {\rm and}\qquad}
\newcommand{\dz}[1]{\frac{\partial  #1}{\partial z}}

\begin{document}

\title[Completeness of inertial modes in an ellipsoid]{Completeness of
Inertial Modes of an Incompressible Inviscid Fluid in a Corotating
Ellipsoid}

\author{George Backus}

\affiliation{
Scripps Institution of Oceanography\\
University of California, San Diego
La Jolla, CA 92093-0225}
\email{gbackus@ucsd.edu}

\author{Michel Rieutord}
\affiliation{
Universit\'e de Toulouse; UPS-OMP; IRAP; Toulouse, France\\
CNRS; IRAP; 14, avenue Edouard Belin, F-31400 Toulouse, France}
\email{Michel.Rieutord@irap.omp.eu}

\date{\today}

\begin{abstract}
Inertial modes are the eigenmodes of contained rotating fluids restored
by the Coriolis force. When the fluid is incompressible, inviscid
and contained in a rigid container, these modes satisfy Poincar\'e's
equation that has the peculiarity of being hyperbolic with boundary
conditions. Inertial modes are therefore solutions of an ill-posed
boundary-value problem. In this paper we investigate the mathematical
side of this problem. We first show that the Poincar\'e problem can
be formulated in the Hilbert space of square-integrable functions,
with no hypothesis on the continuity or the differentiability of
velocity fields. We observe that with this formulation, the Poincar\'e
operator is bounded and self-adjoint and as such, its spectrum is the
union of the point spectrum (the set of eigenvalues) and the continuous
spectrum only. When the fluid volume is an ellipsoid, we show that the
inertial modes form a complete base of polynomial velocity fields for the
square-integrable velocity fields defined over the ellipsoid and meeting
the boundary conditions. If the ellipsoid is axisymmetric then the base
can be identified with the set of Poincar\'e modes, first obtained by
Bryan (1889) \cite{bryan1889}, and completed with the geostrophic modes.
\end{abstract}

\maketitle

\section{Introduction}

Rotation is a ubiquitous feature in stars, planets and satellites. The
dynamics of these objects is profoundly modified when solid body
rotation overwhelmingly dominates all other flows. In this case
residual disturbances that make the flow depart from an exact solid body
rotation are strongly affected by the Coriolis acceleration which
ensures angular momentum conservation of the movements. This is
especially true for the low frequency oscillations of stars or planets.
For these oscillations buoyancy and Coriolis force are the
restoring forces at work. They make gravito-inertial waves possible
\cite[][]{FS82a,DRV99}.

In stars these waves are of strong interest because their detection
and identification allow us to access to both the \BVF\
distribution as well as the local rotation of the fluid. They are of
particular interest in massive stars, where they open a window on the
interface separating the inner convective core and the outer radiative,
and stably stratified, envelope. But these waves are also a key feature
of the response of tidally interacting bodies and therefore of their secular
evolution \cite[][]{OL04,O05,RV10,O14}. On this latter subject several
studies have recently addressed the dynamics of fluid flows driven by
librations, which are common phenomena in planetary satellites
\cite[e.g.][]{SLD13,Zhang_etal12,Zhang_etal13}.

However, the mathematical problem set out by these global oscillations
is far from being fully understood. The reason for that comes from
the very basic boundary value problem that emerges when diffusion and
compressibility effects are neglected: it is ill-posed mathematically
\cite[][]{Green69}. The operator is indeed either of hyperbolic or mixed
type in the spatial coordinates, but the solutions need to match boundary
conditions. As already noted by many authors after the seminal work of
Hadamard \cite{hadamard32}, ill-posed problems are plagued with many
sorts of singularities \cite[e.g.][for a detailed discussion]{RGV01}.

With planetary and stellar applications in mind the oscillations of an
incompressible fluid confined in a rotating sphere or spherical shell have
attracted much attention \cite[][]{HK95,RV97,RGV00,RGV01,RV10,SLD13}. The
oscillating flows in a spherical shell display strong singularities when
viscosity vanishes \cite[][]{RGV01}. The singularities occur because
perturbations obey the spatially hyperbolic Poincar\'e equation
(see Eq~\ref{poincpb} below), and must meet boundary conditions. The
strongest singularities, called wave attractors after the work of Maas \&
Lam \cite[][]{ML95}, result from the reflection of the characteristic
lines (or surfaces) on the boundaries\footnote{Note that on the well-posed
hyperbolic problem -- Cauchy problem -- where initial conditions replace
boundary conditions on the time-coordinate, there is no reflection
towards the past!}. In the two-dimensional problem analog to that of the
spherical shell, characteristic lines are focussing around periodic
orbits (the attractors) \cite[][]{RVG02}. It can be further shown that
no eigenmode can exist when an attractor is present \cite[][]{RGV01}. Of
course viscosity regularizes the solutions, but numerical solutions of
the viscous eigenvalue problem show that actual eigenmodes are strongly
featured by attractors. They appear as thin oscillating shear layers
attached to the attractor.

Surprisingly, when the inner core of the spherical shell is suppressed,
namely the container is a full sphere (or a full ellipsoid)
regular polynomial solutions exist for the inviscid eigenvalue
problem \cite[][]{vantieghem14}. For the sphere and the axisymmetric
ellipsoid, these solutions have long been known since the paper of
Bryan \cite{bryan1889}, which followed the seminal work of Poincar\'e
\cite{Poinc1885} on the equilibrium of rotating fluid masses \cite[but
see also][]{Green69,ZELB01}.

When Greenspan \cite{Green69} reviewed the subject in his monograph
on rotating fluids, he raised the question of the completeness of
the inertial modes in the sphere and the ellipsoid. Indeed, if the
normal modes are complete, then any perturbation can be expanded into a
linear combination of
eigenfunctions. In particular any initial condition can be expanded
and the response flow can be calculated, while perturbations by viscous
or nonlinear effects can be easily dealt with. Except for the work of
Lebovitz \cite{lebovitz89} (see below), Greenspan's question remained
untouched for almost fifty years until the recent works of Cui et
al. \cite{cui_etal14}, who proved completeness for the rotating annular
channel, followed by the one of Ivers et al. \cite{IJW15} who gave the
demonstration for the sphere.

The present work, which has an unusual history (see the end of the paper),
extends the results of Ivers et al. to any ellipsoid.  Importantly,
our demonstration takes another route than the one found by Ivers
et al.\cite{IJW15}. We use a more general formulation of the problem
allowing us to use the tools of functional analysis in the Hilbert space
of square-integrable functions. Since these tools are likely unfamilar to
many fluid dynamicists, we try to make our demonstration as pedagogical
as possible.

The paper is organised as follows. In the next section we first
formulate the Poincar\'e problem, either for forced flows or for free
oscillations. Then, in section 3, we propose another formulation of the
free oscillation
problem that does not assume continuity or differentiability of velocity
fields. Velocity fields are only supposed to be square-integrable. Such
an extension of the space of velocity fields is motivated by three
arguments: first, inviscid fluid may support discontinuous velocity
fields, like the classical vortex sheet \cite[][]{rieutord15}. Second,
singular velocity field can be expected because of the ill-posed nature
of the Poincar\'e problem. Third, and not least, by assuming only
square-integrability of the solution of the problem, we can play in
the Hilbert space of square-integrable functions, and benefit from many
results of spectral theory on bounded, self-adjoint, linear operators. In
section 4, we summarize what we can readily say about this problem
using some of the results of functional analysis, recalling in passing
the needed concepts of spectral analysis. We then
establish a sufficient condition for an operator to own a complete
basis of eigenfunctions. We show that polynomial eigenfunctions can
constitute such a base if the fluid volume is an ellipsoid. This result
was also obtained by Lebovitz \cite{lebovitz89}, but our proof is more direct
and clearly exhibit the special nature of the ellipsoidal boundary. In
section 5, we consider the well-known (since Bryan 1889)
eigenmodes of the rotating spheroid (i.e. the axisymmetric ellipsoid). These
solutions are of polynomial nature and we show (section 6) that they
constitute the expected complete base that has been infered in the
previous section. Notably, we exhibit the set of geostrophic modes that
are associated with the zero-eigenfrequency, and without which inertial
modes would not make a complete base.

The present work is therefore a follow up of the work of Ivers et
al. \cite{IJW15} who obtained a first set of mathematical results when
the problem is restricted to the sphere and when the velocity fields
are supposed to be once-continuously differentiable. The two works
share many common results, but hopefully they complete one another and
offer the broadest view of the Poincar\'e problem. The method proposed
here seems promising enough that one might hope to use it when the fluid
volume is not an ellipsoid.  We have investigated two other shapes,
a cube and a spherical shell, with only negative results. Hence, except
the annular channel \cite[][]{cui_etal14}, we simply do not know whether
any non-ellipsoidal volume has a complete set of eigenvelocities of some
more general form.

\section{Classical Formulation of the Poincar\'e problem}

In the steady, undisturbed reference state, an incompressible non-viscous
fluid with constant density $\rho$ occupies an open bounded set $E$
with boundary $ \partial E $ that has an outward unit normal $ {\hat
{\bf n}} $. Let $\overline{E}$ be the closure\footnote{We recall that
the closure of a metric space $S$ includes the set itself plus all the
limits of converging suites defined on the set $S$. Hence, the set of
real numbers is the closure of the set of rational numbers.}
 of $E$, i.e. $E$ together
with $\partial E$. Both $ \partial E $ and the fluid rotate rigidly
about some given axis with constant angular velocity ${\bf\Omega} $.
Position vectors $ {\bf r} $ are measured relative to an origin chosen
on the axis of rotation.  The body force on the fluid in the rotating
reference frame is independent of time and consists of self-gravity,
externally applied gravity, and centrifugal force.  The pressure in the
fluid is the hydrostatic pressure required to balance these body forces.

In the disturbed state $ \partial E $ is infinitesimally deformed
to $ \partial E_t $ at time $t$, and the infinitesimal
normal velocity of $ \partial E_t $ is $ \beta $.  An extra
infinitesimal time-dependent body force $ {\bf f} $ per 
unit mass acts on the fluid.  In consequence of these forces and
its own history, the fluid has an
infinitesimal velocity $ {\bf v} $ when viewed from the 
rotating frame.  The hydrostatic pressure suffers an
infinitesimal perturbation which it will be convenient to write
as $2 \rho \Omega q$, where $\Omega$ is the magnitude of
$ {\bf\Omega}$ and $q$ is a function of ${\bf r}$ and
the time $t$.  In the rotating reference frame, $ {\bf v} $ and
\it q \rm are governed by the equations

\begin{subequations}
\begin{equation}
{\hat {\bf n}} \,\cdot\, {\bf v} \ =\  \beta \qquad  {\rm on} \quad \partial E,
\label{bc1}
\end{equation}
\begin{equation}
 \nabla \cdot {\bf v} \ =\ 0 \qquad  {\rm in} \quad  E,
\label{eqmass}
\end{equation}
\label{eqbc1}
\end{subequations}

\begin{equation}
\partial_t {\bf v} \ +\ 2\,  {\bf\Omega} \, \times {\bf v}
\ =\  -2 \Omega  \nabla q \ +\  {\bf f} \ \ \ \ \ \ 
{\rm in} \ \ \  E.
\label{eqmom}
\end{equation}
Because $\rho$ is constant, \eq{eqmass} is exact, but \eq{bc1} and
\eq{eqmom} are correct only to first order in the disturbances $ \beta $,
$ {\bf v} $, $q$ and $ {\bf f} $.  For simplicity it will be assumed that
$ \beta $ and $ {\bf f} $ are known for all $t>0$, and that $ {\bf v}
$ is known everywhere at $t\ =\ 0$.  Using this information to find $
{\bf v} $ and $q$ for all $ {\bf r} $ in $E$ and all $t>0$ constitutes
the Poincar\'e forced initial value problem.

It will be convenient to eliminate $\beta $ at the outset.  If
$\beta \neq 0$, let $\theta$ be a solution of the following
Neumann problem (Kellogg, 1953, p246) at each time\nocite{kellogg53}
$t$ :

\begin{subequations}
\begin{equation}
{\hat {\bf n}} \,\cdot\, \nabla \theta \ =\  \beta \qquad       {\rm on}
\ \ \  \partial E,\label{neum}
\end{equation}
\begin{equation}
\nabla^2 \, \theta \ =\  0 \qquad       {\rm in} \ \ \  E.
\end{equation}

The solubility conditions for this Neumann problem are that $\partial E$
be sufficiently smooth (for example, $\hat{\bf n}$ may vary continuously
on $\partial E$) and that
\begin{equation}
\int_{ \partial E } \,dA\  \beta \ =\ 0, \label{intsur}
\end{equation}
\end{subequations}
a condition whose fulfillment is assured by \eq{eqbc1}.  Given
\eq{intsur}, the solution $ \theta $ of (\ref{neum},b) is determined at
each $t$ up to an unknown additive function of $t$,
and $  \nabla \theta $ is uniquely determined for all $t$.
If we define 

\begin{subequations}
\begin{equation}
{\bf v}^\prime \ =\  {\bf v} \ -\   \nabla \theta
\end{equation}
then $ {\bf v}^\prime $ satisfies equations \eq{eqbc1} and \eq{eqmom}
with $ \beta $ replaced by 0, with $q$ replaced by 

\begin{equation}
q^\prime \ =\ q \ +\  (2 \Omega )^{-1} \  \partial_t \theta
\end{equation}
and with $ {\bf f} $ replaced by

\begin{equation}
{\bf f}^\prime \ =\  {\bf f}  \ -\  2  {\bf\Omega} \ 
\times \   \nabla \theta .
\end{equation}
\label{eq24}
\end{subequations}
Henceforth we drop the primes and take $ \beta \ =\  0$ in \eq{bc1}.

To find the normal modes we set $ {\bf f} \ =\   {\bf 0} $ in \eq{eqmom}
and look for solutions of \eq{eqbc1} and \eq{eqmom} whose time
dependence is

\begin{subequations}
\begin{equation}
{\bf v} ( {\bf r} ,t) \ =\  {\bf v} 
( {\bf r} ,0)\ e^{{2i} \Omega \lambda t}
\end{equation}
\begin{equation}
q( {\bf r} ,t) \ =\  q( {\bf r} ,0)\ 
e^{{2i} \Omega \lambda t}
\end{equation}
\end{subequations}
where $\lambda$ is an unknown complex constant.  In studying the normal
modes we will abbreviate $ {\bf v} ( {\bf r} ,0)$
and  $q( {\bf r} ,0)$ as $ {\bf v} ({\bf
r} )$ and $q( {\bf r} )$ or simply as $ {\bf v} $
and $q$. In these circumstances, \eq{eqbc1} and \eq{eqmom} are replaced by

\begin{subequations}
\begin{equation}
{\hat {\bf n}} \,\cdot\, {\bf v} \ =\  0 \qquad  {\rm on} \quad
\partial E, \label{bc2}
\end{equation}
\begin{equation}
 \nabla \,\cdot\, {\bf v} \ =\ 0 \qquad  {\rm in} \quad
E,\label{eqmass2}
\end{equation}
\label{bcmass2}
\end{subequations}

\begin{equation}
- \lambda {\bf v} \ +\ i\,  {\hat{\bf\Omega}} \, \times {\bf v}
\ =\ -i\,  \nabla q  \qquad  {\rm in} \quad  E,\label{inert}
\end{equation}
where $  {\hat{\bf\Omega}} \ =\   {\bf\Omega} / \Omega $, the
unit vector in the direction of $  {\bf\Omega} $.  

Kudlick (1966) \cite{kudlick66} and Greenspan (1968) show that when $ {\bf v} $
and $q$ are smooth enough to permit some differentiation then $ \lambda
$ cannot be $+1$ or $-1$.  We will treat the geostrophic case ($ \lambda\
=\ 0$) later, so for the moment we assume that $ \lambda $ is not 0, +1
or $-1$.  Then (Greenspan, 1968, p.51) equation \eq{inert} can be solved for $
{\bf v} $ in terms of $  \nabla q$ to produce

\begin{equation}
\lambda (1\,-\, \lambda^2 )\, {\bf v} \ =\ -i\,
\lambda^2 \,  \nabla q \ +\  \lambda \,  {\hat{\bf\Omega}} \,
\times \,  \nabla q \ +\ i\, 
\hat{\bf\Omega} \;  {\hat{\bf\Omega}}\cdot\nabla q\; .\label{vvp}
\end{equation}
Substituting \eq{vvp} in \eq{bc2} gives

\begin{subequations}
\begin{equation}
\lambda^2 \, {\hat {\bf n}} \,\cdot\,  \nabla q \ +\  i \lambda
( {\hat {\bf n}} \, \times \,  {\hat{\bf\Omega}} )\,\cdot\, 
 \nabla q \ =\  ( {\hat {\bf n}} \,\cdot\,  {\hat{\bf\Omega}} )\,
(  {\hat{\bf\Omega}} \,\cdot\,  \nabla q) \ \ \ \ \ \  {\rm on} \ \ \  
\partial E.\label{bcpoinc}
\end{equation}
Substituting \eq{vvp} in \eq{eqmass2} gives 

\begin{equation}
(  {\hat{\bf\Omega}} \,\cdot\,  \nabla )^2 \,q\ =\  \lambda^2
\, \nabla^2 \,q \qquad  {\rm in} \quad  E.\label{poinc}
\end{equation}
\label{poincpb}
\end{subequations}
Equation \eq{poinc} is the classical Poincar\'e equation for the pressure
disturbance $q$, and \eq{bcpoinc} is the boundary condition appropriate
to the Poincar\'e problem, in which $ \partial E $ rotates rigidly.
Given an eigenfunction $q$ and its eigenvalue $ \lambda $ in
\eq{poincpb}, the
corresponding $ {\bf v} $ is recovered from \eq{vvp}.  Greenspan (1965)
\cite{greenspan65}
shows that when $ {\bf v} $ is sufficiently differentiable then $
\lambda $ must be real and between $-1$ and 1.  The resulting hyperbolic
character of \eq{poinc} for the normal modes has led to the suspicion that
there might be pathological elements in the boundary value problem
\eq{poincpb} \cite[][]{SR69}.

\section{Admitting non-differentiable velocity fields}\label{sect3}

\subsection{Introduction}

Inviscid incompressible fluids admit discontinuous velocity fields
provided discontinuities are parallel to the field so as to fulfill
mass conservation. Hence, eigenvalues may be associated with
non-differentiable velocity fields. In view of the ill-posed nature of
the Poincar\'e problem, the possibility of such eigen-velocities cannot be
excluded. In this section we therefore reformulate the eigenvalue
problem \eq{bcmass2}-\eq{inert} in order to include non-differentiable
velocity fields.

Under suitable smoothness assumptions Greenspan (1964, 1965)
\cite{greenspan64,greenspan65} shows that,
whatever the shape of the fluid volume $E$, all eigenvalues $\lambda$
of \eq{bcmass2} and \eq{inert} are real and lie in the interval
$-1<\lambda<1$.  That author also shows that eigenvelocities $\, {\bf
v}_1$ and $\, {\bf v}_2$ belonging to different eigenvalues $\lambda_1$
and $\lambda_2$ are orthogonal in the sense that $\langle \  {\bf v}_1
\,|\, {\bf v}_2 \, \rangle \ =\  0$, where the inner product is defined as
\begin{equation}
\langle \  {\bf v}_1 \,|\, {\bf v}_2 
\, \rangle \ =\ 
|E|^{-1} \ \int_E \, dV ( {\bf r} )\ {\bf v}_1 ( {\bf r} )^* \cdot\, {\bf v}_2 
( {\bf r} ) \;. \label{innprod}
\end{equation}
Here $|E|$ is the volume of the region $E$, and ${\bf v}_1 ( {\bf r}
)^*$ is the complex conjugate of ${\bf v}_1 ( {\bf r} )$.

All this suggests that the eigenvalues $\lambda$ are the eigenvalues of
some bounded, self-adjoint linear operator $L$ on the complex Hilbert
space $  \ubPi $ consisting of all Lebesgue square-integrable
complex vector fields ${\bf v}$ on $E$.  We recall that
square-integrability just means that the total kinetic energy of the
flow exists. For such velocity fields,
we can define their norm by

\begin{equation}
 \|  {\bf v}  \|  \ =\ \langle \  {\bf v} \,|\, {\bf v}
\, \rangle \,^{{1\over 2}} \, .
\label{eq32}\end{equation}
Now, to find the appropriate operator $L \,:\,  \ubPi \ \to\   \ubPi$, we
must interpret \eq{bcmass2} and \eq{inert} when ${\bf v}$ is merely
square-integrable and not differentiable or even continuous.

\subsection{Mass conservation for $\calL^2$-velocity fields}

For velocity fields ${\bf v}$ that are merely
square-integrable and not differentiable or even continuous 
$\nabla \cdot\, {\bf v}$ is not well-defined
in $E$, and $\hat{\bf n} \,\cdot\, {\bf v}$ is not well-defined\footnote{A
square-integrable vector field may indeed not be defined on $\partial E$,
namely on a set of volume measure 0 in $E$.} on $\partial E$.  We begin
by trying to avoid this difficulty.

The game will be to define subspaces of the general Hilbert space $\ubPi$
that includes all the square-integrable complex vector fields ${\bf v}$
defined on $E$. To ease reading, we shall use underlined symbols to
denote a space (of functions usually). It'll be boldface if the space is
a space of vectorial functions. Thus, we first introduce $\uPi^{\infty}$
and $ \ubPi^{\infty}$ that are respectively the spaces of all infinitely
differentiable complex scalar and vector fields on $\overline{E}$,
the closure of $E$.  Define

\begin{subequations}
\begin{equation}
 \ubGa^{\infty}  :=   \nabla \Pi^\infty\, .
\end{equation}
That is, $ \ubGa \,^{\infty}$ consists of all vector 
fields ${\bf u}$ which can be written
\begin{equation}
{\bf u} \ =\   \nabla \, \phi \label{potflow}
\end{equation}
\end{subequations}
for some $\phi$ in $\Pi^{\infty}$.  Then clearly $ \ubGa^{\infty} \,
\subseteq  \,  \ubPi$, but $ \ubGa \,^{\infty}$ is not closed in $ \ubPi$
under the norm \eq{eq32}. Indeed, we can easily construct a suite of
infinitly differentiable function  that converges to a discontinuous
function. Therefore, we consider its closure, $ \ubGa$:

\begin{equation}
 \ubGa  :=  \overline{\ubGa^\infty} \,.
\end{equation}
According to this definition, a vector field ${\bf u}$ on $E$ belongs
to $ \ubGa$ if and only if it is square-integrable on $E$ and there is
a sequence $\phi_1 ,\  \phi_2 ,\, ...$ in $\uPi^{\infty}$ such that

\begin{equation}
\lim_{n\to\infty}  \|{\bf u} -  \nabla\phi_n \|  =  0 \,.
\end{equation}
In particular, $\ubGa$ includes all fields ${\bf u}$ of form \eq{potflow}
with $\phi$ continuously differentiable on $\overline{E}$.

Let us now introduce $\ubLam$, the
orthogonal complement of $\ubGa$ in $\ubPi$.
Thus $\ubLam$ consists of all vector fields
${\bf w}$ square-integrable on $E$ and such that
$\langle\, {\bf u} | {\bf w} \,\rangle = 0$
for every ${\bf u} $ in $\ubGa$.
In particular, ${\bf w} \in \ubLam$ implies that

\begin{equation}
|E|^{-1} \  \int_E \, dV ( {\bf r} )\ (\nabla\phi^*)\cdot{\bf w} \ =\  0
\label{eq36}\end{equation}
for every $\phi$ in $\uPi^{\infty}$.  Conversely, since the orthogonal
complement of a set is also the orthogonal complement of its closure, if
${\bf w}$ is square-integrable on $E$ and \eq{eq36} is true for every $\phi$
in $\uPi^{\infty}$, then ${\bf w} \in \ubLam$.

Now suppose ${\bf w} \in\ubLam \cap \ubPi^{\infty}$.
Then, Gauss's theorem permits \eq{eq36} to be rewritten as

\begin{equation}
\int_{\partial E} dA \phi^*(\hat{\bf n}\cdot{\bf w}) -
\int_E dV \phi^*(\nabla\cdot{\bf w})= 0\quad \forall\phi\in\uPi^\infty \,.
\label{eq37}
\end{equation}
By the Weierstrass approximation theorem (Courant and Hilbert, 1953, p65)
every $\phi$ continuous on $\overline E$ can be approximated uniformly
and with arbitrary accuracy by polynomials.  Therefore \eq{eq37} holds
for all $\phi$ continuous on $\overline E$. Then a well-known argument
leads to the conclusion that $ \nabla \,\cdot\, {\bf w} \,=\, 0$ in
$E$ and ${\hat {\bf n}} \,\cdot\, {\bf w} \,=\, 0$ on $\partial E$.
Therefore, the demand

\begin{equation}
{\bf v} \in \ubLam\label{eq38}
\end{equation}
\par\noindent
is the appropriate generalization of \eq{bcmass2} to square-integrable
vector fields which are not differentiable.

\subsection{$\ubLam$ and piecewise continuously differentiable fields}

Before going any further, it is worth viewing \eq{eq38} from a physical
point of view. $\ubLam$ is indeed a very large space that includes,
among other fields, unbounded vector fields that are not physically acceptable.

We know that the local equation $\nabla\cdot\vv=0$ is
equivalent to the integral condition

\beq \intsur \vv\cdot\dS = 0\qquad \forall \; S\in\overline{E}\eeqn{massflux}
when $\vv$ is differentiable. It says that for any closed surface $S$,
contained in $\overline{E}$, the mass-flux across this surface is zero
(for a fluid of constant density). We shall see now that being a
piecewise continuous vector field in $\ubLam$ is equivalent to
\eq{massflux} being satisfied.

Let us first observe that if $\vv$ is a once-continuously differentiable that verifies
\eq{bcmass2}, then for any $\phi$, a once-continuously
differentiable function of $\uPi^\infty$, we have

\beq \int_E\na\cdot(\phi\vv)\; dV = \int_{\partial E}\phi\,\vv\cdot\dS = 0\eeq
so that

\beq \int_E(\phi\na\cdot\vv + \vv\cdot\na\phi)\;dV = \int_E\vv\cdot\na\phi\;dV=0
\eeqn{ortho}
which shows that such $\vv$-fields are members of $\ubLam$. Now,
Eq.~\ref{ortho} implies \eq{bcmass2} by the reasoning following \eq{eq37}.

However, we can also be slightly less restrictive on $\vv$ and just assume a
piecewise continuous field. Then we can show that for such fields
$\ubLam$-membership is equivalent to \eq{massflux}.

If $\ubLam$-membership \eq{eq38} is true, then for any real $\phi$,
once-continuously differentiable function of $\uPi^\infty$, we have

\beq \int_E \vv\cdot\na\phi\; dV =0\eeqn{cond1}
However, $\na\phi$ is a vector that is always orthogonal to any
iso-$\phi$ surface. Since \eq{cond1} is true for any $\phi$, for a given
surface $S$ we can design a $\phi$ that is constant inside $S$ and
outside $S+\delta S$. $S+\delta S$ is the same as $S$ but dilated by a
small increment $\delta\ell$. In between the two surfaces $\phi$ is
chosen to increase linearly by the same amount so that $\|\na\phi\|$
is the same everywhere on the surface. Hence, for this given $\phi$,
\eq{cond1} implies that

\beq \int_S \vv\cdot\vn \|\na\phi\| dS\delta\ell = 0\eeqn{massfluxb}
where $\vn$ is the unit vector $\na\phi/\|\na\phi\|$ normal to the
surface. Since $\phi$ is chosen such that $\delta\ell$ and $\|\na\phi\|$
are constant, we can simplify \eq{massfluxb} and get \eq{massflux}. We
note that since $\phi$ is any function of $\uPi^\infty$ we can construct
suites of functions whose limit can fit any closed surface, even with
sharp angles.  Hence, all piecewise continuous members
of $\ubLam$ satisfy mass conservation expressed in \eq{massflux}.

Now we would like to know if $\ubLam$ contains all the mass-conserving
velocity fields. Let us therefore show that a piecewise continuous
field verifying  \eq{massflux}
is necessarily in $\ubLam$. For that we prove that if this is not the
case then we get a contradiction. We thus consider a real velocity field
that verifies \eq{massflux} but that does not belong to $\ubLam$. Hence,
there exists a scalar field $\phi\in\Pi^\infty$ defined over the full
volume $E$ such that

\beq \int_E \vv\cdot\na\phi\; dV \neq0\eeqn{cond2}
To make the reasoning easier to follow, we shall assume in addition that
$\phi$ is a monotonic function over $E$. If this is not the case then $E$
can be split into sub-volumes where it is monotonic, and the following
reasoning applies to each sub-volume.

Since $\phi$ is defined over E, the equation

\[ \phi(x,y,z)=\phi(x_0,y_0,z_0)=\phi_0\]
defines a surface which contains the point $(x_0,y_0,z_0)\in E$. Since
$\vv$ is a mass-conserving velocity field, \eq{massflux} is true for any
closed surface, in particular for the surface $\phi=\phi_0$. If this
surface is not closed, then it is completed by the needed part of
$\partial E$. Thus we can write

\[ \int_{\phi=\phi_0} \vv\cdot\dS=0 = \int_{\phi=\phi_0}
\vv\cdot\na\phi\frac{dS}{\|\na\phi\|}\]
Since $\phi$ is a function defined all over $E$, let $\phi_m$ and
$\phi_M$ be the minimum and maximum value reached by $\phi$ in $E$, then

\[ \int_{\phi_m}^{\phi_M} \int_{\phi=\phi_0}
\vv\cdot\na\phi\frac{dS}{\|\na\phi\|}d\phi_0 = 0\]
However, $d\phi_0/\|\na\phi\|$ is the differential length element
orthogonal to the surface, hence $dSd\phi_0/\|\na\phi\|$ is just the
volume element. When $\phi$ scans the interval $[\phi_m,\phi_M]$ the
surface $\phi=\phi_0$ scans the volume $E$. We thus find that 

\beq \int_E \vv\cdot\na\phi\; dV =0\eeqn{cond3}
in contradiction with \eq{cond2}. 

To conclude, we see that all piecewise continuous
velocity fields of $\ubLam$ satisfy mass conservation in its integral
formulation \eq{massflux} and reciprocally.
However, let us stress again that $\ubLam$ is a much wider space that
includes vector fields for which \eq{massflux} or $\na\cdot\vv$ may not make
sense. Its vector fields are just square-integrable and verify \eq{eq36},
which will be sufficient for our purpose.

\subsection{The momentum equation}

We need a similar generalization of the equation of momentum. \eq{inert}
has no derivative in the velocity field, so the question is just a
matter of how to reduce the functional space $\ubPi$ to $\ubLam$.

Since $\ubGa$ is closed, and $ \ubLam$ is its orthogonal complement in $
\ubPi$, therefore

\begin{equation}
 \ubPi \ =\   \ubGa \ \oplus\   \ubLam \,.
\end{equation}
\par\noindent
That is, every ${\bf v}$ in $ \ubPi$ can be written in
the form ${\bf v} \,=\, {\bf u} \,+\, {\bf w}$ with
${\bf u} \,\in \,  \ubGa$ and 
${\bf w} \,\in\,  \ubLam$, and
$\langle \  {\bf u} \,|\, {\bf w} \, \rangle \,=\, 0$.
The foregoing definitions are very similar to the decomposition of the
classical vector space into two orthogonal subspaces (like a plane and a
line in $\bbbr^3$). In the following we just identify the projection
operators on the subspaces.

The orthogonality of the subspaces $\ubLam$ and $\ubGa$ means that ${\bf
u}$ and ${\bf w}$ are uniquely determined by ${\bf v}$, so that it is
possible to define two functions, $\Gamma \,:\,  \ubPi \,\to\,  \ubGa$
and $\Lambda \,:\,  \ubPi \,\to\,  \ubLam$, as follows: for any ${\bf v}$
in $ \ubPi$

\begin{subequations}
\begin{equation}
{\bf v} \ =\  \Gamma ( {\bf v} ) \ +\  \Lambda ( {\bf v} )
\end{equation}
where
\begin{equation}
\Gamma ( {\bf v} ) \ \in\   \ubGa \,, \ \ \ \ \ \ 
\Lambda ( {\bf v} ) \ \in\   \ubLam \,.
\end{equation}
\end{subequations}
\par\noindent
From the uniqueness of $\Gamma ( {\bf v} )$ and 
$\Lambda ( {\bf v} )$ it follows that $\Gamma$ and $\Lambda$
are linear, and since
$\langle \  \Gamma  {\bf v}  \,|\, \Lambda  {\bf v}  \, 
\rangle \,=\,0$ it follows that $ \|  {\bf v}  \|^2 \,=\, 
 \|  \Gamma  {\bf v}   \|^2 \,+\, 
 \|  \Lambda  {\bf v}   \|^2$.
Thus $ \|  \Gamma {\bf v}  \|  \,\le\,  \|  {\bf v} ||$ and
$ \|  \Lambda {\bf v}  \|  \,\le\,  \|  {\bf v} ||$.
The functions $\Gamma$ and $\Lambda$ are the orthogonal
projectors of $ \ubPi$ onto $ \ubGa$ and
$ \ubLam$.
They are bounded linear operators on $ \ubPi$ with
the following properties (see Lorch, 1962, p72):\nocite{lorch62}
\begin{subequations}
\begin{equation}
{\rm I} \,_{{} \ubPi} \ =\  \Gamma \ +\  \Lambda
\label{eq311a}\end{equation}
\begin{equation}
\Gamma^2 \ =\  \Gamma \,, \ \ \ \ \Lambda^2 \ =\  \Lambda
\label{eq311b}\end{equation}
\begin{equation}
\Gamma \Lambda \ =\  \Lambda \Gamma \ =\  0
\label{eq311c}\end{equation}
\begin{equation}
 \|  \Gamma  \|  \ =\   \|  \Lambda  \|  \ =\  1
\label{eq311d}\end{equation}
\begin{equation}
\Gamma^* \ =\  \Gamma \,, \qquad   \Lambda^* \ =\  \Lambda
\label{eq311e}\end{equation}
\begin{equation}
\Gamma  \ubPi \ =\   \ubGa \,, 
\qquad   \Lambda  \ubPi \ =\   \ubLam \,.
\label{eq311f}\end{equation}
\end{subequations}
\par\noindent
Here $ {\rm I} \,_{{} \ubPi} $ is the identity operator on $  \ubPi $,
and for any linear operator $F$ on $\ubPi$, $\| F\|$ is its norm, namely

\[ \| F\| = \sup \{\| F {\bf v} \| :  \| {\bf v} \| = 1  \} \;,\]
and $F^*$ is its adjoint.  The three statements ${\bf u} \,\in\,  \ubGa$,
$\Gamma {\bf u} \,=\, {\bf u}$ and $\Lambda {\bf u} \,=\,  {\bf 0}$
are equivalent, as are the three statements ${\bf w} \,\in\,  \ubLam$,
$\Lambda {\bf w} \,=\, {\bf w}$ and $\Gamma {\bf w} \,=\,  {\bf 0}$.

When ${\bf v}$ and $\Gamma {\bf v}$ belong to $ \ubPi^{\infty}$, it
is easy to compute $\Gamma {\bf v}$ and $\Lambda {\bf v} \,=\, {\bf v}
\,-\, \Gamma {\bf v}$ as follows.  Let ${\bf u} \,=\, \Gamma {\bf v}$
and ${\bf w} \,=\, \Lambda {\bf v}$.  Then ${\bf w} \,\in\,\ubLam
\,\cap\,  \ubPi^{\infty}$, so ${\bf w}$ satisfies \eq{bcmass2}.  Also,
${\bf u} \,=\,  \nabla \phi$ for some $\phi \,\in\, \Pi^{\infty}$, so

\begin{equation}
{\bf v} \ =\   \nabla \phi \ +\  {\bf w}
\label{eq312}
\end{equation}
Then, because ${\bf w}$ satisfies \eq{bcmass2},

\begin{subequations}
\begin{equation}
\nabla^2 \phi \ =\   \nabla \,\cdot\, {\bf v} \ \ \ \ 
{\rm in} \ \ \ \ E \,;
\end{equation}
\begin{equation}
\hat{\bf n} \,\cdot\,  \nabla \phi \ =\  \hat{\bf n}
\,\cdot\, {\bf v} \ \ \ \ {\rm on} \ \ \ \  \partial E \,.
\end{equation}
\label{eq313}
\end{subequations}
Since ${\bf v}$ is given, equations \eq{eq313} constitute an
interior Neumann problem for $\phi$ (Kellogg, 1953, p246).
The solubility condition for this problem is

\[ \int_E \,dV\,  (\nabla \,\cdot \, {\bf v} ) \ =\ 
\int_{{\partial} E} \,dA\, ( \hat{\bf n} \,\cdot\, {\bf v} )\;, \]
a condition whose validity is guaranteed by Gauss's theorem.  Therefore,
\eq{eq313} has a solution $\phi$, unique up to an additive constant.
Then $\Gamma {\bf v} \,=\, {\bf u} \,=\,  \nabla \phi$ is uniquely
determined by \eq{eq313}, and $\Lambda {\bf v} $ is the $ {\bf w} $ of
\eq{eq312}. {We note that \eq{eq312} is the weak formulation of the
classical Helmholtz decomposition of three-dimensional vector fields
\cite[see][for a mathematical discussion of divergence-free vector fields
in three-dimensional domains]{amrouche_etal98}.

With the foregoing preliminaries we now return to the momentum equation
\eq{inert}. Define the linear operator $R:\,  \ubPi \,\to\,  \ubPi$
by requiring that for any ${\bf v}$ in $ \ubPi$

\begin{equation}
R {\bf v} \ =\  i \, \hat { {\bf\Omega}} \ \times\  {\bf v} \,.
\label{eq314}
\end{equation}
Then \eq{inert} can be written

\begin{subequations}
\begin{equation}
-\, \lambda {\bf v} \ +\  R {\bf v} \ =\ -i\,  \nabla q \,.
\label{eq315a}\end{equation}
Suppose for the moment that $q \,\in\, \Pi^{\infty}$.
Then $ \nabla q \,\in\,  \ubGa \,^{\infty}$, so
$\Lambda \,  \nabla q \,=\, 0$.
Thus if we apply $\Lambda$ to \eq{eq315a} we obtain
\begin{equation}
-\lambda \Lambda {\bf v} \ +\  \Lambda R {\bf v} \ =\  {\bf 0}\,.
\label{eq315b}\end{equation}
\end{subequations}
But this is an equation which makes sense even if ${\bf v}$ is
merely square-integrable, while if ${\bf v} \,\in\,  \ubPi^{\infty}$
then \eq{eq315b} implies \eq{eq315a} for some $q$.  Thus \eq{eq315b}
generalizes \eq{inert} to all square-integrable ${\bf v}$.

Equation \eq{eq315b} can be further simplified, since the 
eigensolution ${\bf v}$ must also satisfy \eq{eq38},
the generalization of \eq{bcmass2}.  As already noted, \eq{eq38} is 
equivalent to  ${\bf v} \,=\, \Lambda {\bf v}$, 
and this permits rewriting \eq{eq315b} as
\begin{subequations}
\begin{equation}
L {\bf v} \ =\  \lambda {\bf v}
\label{eq316a}\end{equation}
where
\begin{equation}
L \ =\  \Lambda \,R \Lambda \,.
\label{eq316b}\end{equation}
\end{subequations}
The operator $L$ is defined on the whole space $\ubPi$, but $L\ubLam \,
\subseteq \, \ubLam$ and $L\ubGa \,=\, \underline{\{\bf 0\}}$.
Hence the only interesting part of $L$ is actually $L |\, \ubLam$, the
restriction of $L$ to $ \ubLam$.

The Poincar\'e problem \eq{bcmass2}, \eq{inert} is now generalized to
square-integrable but possibly nondifferentiable velocity fields ${\bf
v}$. The pair ${\bf v} ,\  \lambda$ solves this generalized Poincar\'e
problem if ${\bf v}$ is an eigenvector and $\lambda$ the corresponding
eigenvalue of the linear operator $L|\, \ubLam$ on the Hilbert space
$ \ubLam$.

Further study of $L$ depends on the observations that

\begin{subequations}
\begin{equation}
 \|  L  \|  \ \le\  1
\label{eq317a}\end{equation}
and
\begin{equation}
L^* \ =\  L \,.
\label{eq317b}\end{equation}
\label{eq317}
\end{subequations}
To prove \eq{eq317a} note from \eq{eq316b} that
$\| L \| \,\le\, \|\Lambda\| \| R\|\|\Lambda\|$.
By \eq{eq311d} therefore $ \|  L  \|  \,\le\,  \|  R ||$.
But since $|\hat{\bf\Omega}| = 1$,
$|R {\bf v} | \,\le\, | {\bf v} |$, and hence
$ \|  R {\bf v}  \|  \,\le\,  \|  {\bf v} ||$.  Thus

\begin{subequations}
\begin{equation}
 \| R \|  \ \le\  1
\label{eq318a}\end{equation}
and \eq{eq317a} follows.  To prove \eq{eq317b}, note that for bounded
linear operators $F,\,G$ on $ \ubPi$ one has $(FG)^* \,=\, G^*F^*$.  Thus,
from \eq{eq316b}, $L^* \,=\, \Lambda^* \,R^*\,\Lambda^*$.  Then from
\eq{eq311e}, $L^* \,=\, \Lambda \,R^* \, \Lambda$, and \eq{eq317b}
will follow if we can prove that

\begin{equation}
R^* \ =\  R \,.
\label{eq318b}\end{equation}
\end{subequations}
This last is simply the assertion that for any ${\bf v}_1$, ${\bf v}_2$ in $ \ubPi$, 

\[ \langle \  {\bf v}_1 \,|\, i\,\hat { {\bf\Omega}} 
\,\times\, {\bf v}_2  \, \rangle \ =\ \langle \  i\, 
\hat { {\bf\Omega}} \,\times\, {\bf v}_1 \,|\, 
{\bf v}_2 \, \rangle \,, \]
a fact evident from \eq{innprod}.

In what follows, $L|\,  \ubLam$ will usually be abbreviated as $L$ when
no confusion can result.  Properties \eq{eq317} of $L$ assure that all
its eigenvalues $\lambda$ are real and lie in the interval $-1 \,\le\,
\lambda \,\le\, 1$.  Because $L$ is self-adjoint, a well-known argument
(e.g., Lorch, 1962, p112) shows that if $L \, {\bf v}_1 \ =\  \lambda_1
\, {\bf v}_1$ and $L \, {\bf v}_2 \ =\  \lambda_2 \, {\bf v}_2$ and
$\lambda_1 \neq \lambda_2$ then $\langle \  {\bf v}_1 \,|\, {\bf v}_2 \,
\rangle \,=\, 0$.

Thus we generalized to square-integrable ${\bf v}$ the results
obtained by Greenspan (1964, 1965) and Kudlick (1966) for continuously
differentiable ${\bf v}$, with one exception: Kudlick (Greenspan, 1968,
p61) shows that for continuously differentiable ${\bf v}$, $\lambda =
\pm 1$ are not eigenvalues.  In fact the numbers $ \lambda = \pm 1$
can be excluded from the eigenvalue spectrum for any ${\bf v}$ which
is merely square-integrable, and for any volume. We give the complete
proof in appendix.  For triaxial ellipsoids, $ \lambda \neq \pm 1$
also follows from Lebovitz's (1989) result that all eigenfunctions in
an ellipsoid are polynomials, and thus smooth enough to admit Kudlick's
proof.\nocite{lebovitz89}

\def\pdt{ \partial_t }

\def\Kn{ \underline{\bf K} \,_n }
\def\Hn{ \ubH \,_n }
\def\pd{ \partial }
\def\pdt{ \partial_t }
\def\pdx{ \partial_x }
\def\pdy{ \partial_y }
\def\pdz{ \partial_z }
\def\nomem{ \ \it\hbox{\z\in\h'.15m'\v'.07m'\(sl\v'-.07m'\h'-.15m'} }

\section{Completeness of the eigenfunctions for a triaxial ellipsoid}
\label{triaxal}

\subsection{Introduction}

Generalizing the Poincar\'e problem to square-integrable velocity fields
is useful not only because such fields are needed to describe flows of inviscid
fluids, but also because they make available the spectral theory for bounded,
self-adjoint linear operators in Hilbert space.

Let us briefly summarize what spectral theory tells us about $L$
(i.e. $L|\,\ubLam$) which we know to be a linear self-adjoint bounded
operator defined over a Hilbert space. First this operator is normal
as it (obviously) commutes with its adjoint: $LL^*=L^*L$. Then, for
any nonzero bounded linear operator $F$ on a Hilbert space $\ubH$, the
spectrum $\sigma(F)$ of $F$ is the set of all complex numbers $\lambda$
such that $F-\lambda {\rm I}$ fails to have a bounded linear inverse.
The spectrum is always a non-empty, closed subset of the complex plane
(Lorch, 1962, pp89 \& 94).  If $F$ is bounded, then $| \lambda | \,\le\
\| F\|$ for every $\lambda$ in $\sigma \,(\,F)$ (Lorch, p109).  If $F$
is self-adjoint, then $\sigma \,(\,F)$ is a subset of the real axis
(Lorch, p71).

The spectrum can be divided into three parts known as the point
spectrum (the eigenvalues), the continuous spectrum and the residual
spectrum. These three sets are disjoint and in our case they are
subsets of the real axis interval $[-1,1]$ since $\|L\|\leq 1$. For a
self-adjoint operator, it may be proved that the residual spectrum is
empty \cite[e.g. theorem 9.2-4 in][]{kreyszig78}.  Hence, for our problem
we are just left with the continuous and eigenvalue spectra. In this case,
a complex number $\lambda$ can qualify for membership in $\sigma(F)$ in
two ways: first, there may be a nonzero ${\bf h}$ in $\ubH$ such that
$(F- \lambda {\rm I}){\bf h} = {\bf 0}$; that is, $\lambda$ may be an
eigenvalue of $F$ (its eigenvector being ${\bf h} $). In other words,
when $\lambda$ is in the point spectrum of $F$, $(F- \lambda {\rm I})$
is not injective.  Second, $\lambda$ may be such that $(F- \lambda {\rm
I})^{-1}$ exists but $(F- \lambda {\rm I})$ is not surjective. In other
words, $(F- \lambda {\rm I})(\ubH) \neq \ubH$ but $\overline{(F- \lambda
{\rm I})(\ubH)} = \ubH$ or the image of $(F- \lambda {\rm I})$ is dense
in $\ubH$. In this case $\lambda$ belongs to the continuous spectrum.

Interestingly, another subdivision of the spectrum has been introduced by
mathematicians (e.g. \citealt{halmos51}, p51 or \citealt{furuta01},
p81). This other division is between the approximate point spectrum
and the compression spectrum. Unlike the preceding subsets of the
spectrum, these two subsets are not disjoint. When $\lambda$ is in
the approximate point spectrum $(\,F\,-\, \lambda {\rm I)\,} {\bf h}$
may be nonzero whenever ${\bf h} \neq   {\bf 0} \,$, but there may be a
sequence ${\bf h}_1 \,,\ {\bf h}_2 \,,\ .\,.\,.$ in $\ubH$ such that $
\|  \, {\bf h}_n \|  \ =\ 1$ and $\lim_{{n} \to \infty} \, \|  \,(F\,-\,
\lambda {\rm I)\,} {\bf h}_n  \|  \ =\  0$.  In this case, $(\,F -\,
\lambda {\rm I)}^{-1}$ is a linear mapping well-defined on the range of
$F-\, \lambda {\rm I}$, but it is not a bounded operator and hence has
no linear extension to all of $\ubH$ (Lorch, p44). To be complete the
compression spectrum is the set
\[\sigma_{\rm comp}(F) = \{\lambda\in\bbbc
| \overline{{\rm Range}(F-\lambda{\rm I})}\subsetneq \ubH \}\; ,\]
hence a subset of
the continuous spectrum. However, we learn from \cite[][]{furuta01}
(\S2.4, theorem 12) that for a normal operator the spectrum is identical
to the approximate point spectrum. Applied to the Poincar\'e problem in
the spheroid, which admits a set of eigenvalue dense in [-1,1], we may
identify this interval with the approximate point spectrum and real
numbers that are not eigenvalues are in the continuous spectrum. Of
course, no eigenvectors are associated with members of the continuous
spectrum.

\subsection{A preliminary step}

How do we prove that a bounded, self-adjoint linear operator $F\,:\,
\ubH \,\to\, \ubH$ has a complete set of orthonormal eigenvectors, i.e. a
collection of orthonormal eigenvectors which constitutes an orthonormal
basis for the Hilbert space $\ubH \,$?  One method is to find an infinite
sequence of subspaces of $\ubH \,$, say $ \ubH \,_1  ,\   \ubH \,_2  ,\
\ubH \,_3  ,\  ...$, with these properties:

\begin{subequations}
\begin{equation}
{\rm dim} \  \Hn \ < \infty
\label{eq51a}\end{equation}
\begin{equation}
\Hn \  \subseteq  \  \ubH \,_{n+1}
\label{eq51b}\end{equation}
\begin{equation}
\ubH \ =\  \overline{\, \cup_{n=1}^{\infty} \, \Hn }
\label{eq51c}\end{equation}
\begin{equation}
F \Hn \  \subseteq  \  \Hn \,.
\label{eq51d}\end{equation}
\label{eq51}
\end{subequations}
We claim that whenever such a sequence of subspaces exists, $F$ has
a complete set of orthonormal eigenvectors in $ \ubH $.

To prove this claim, let $\underline{\bf K} \,_1 \ =\   \ubH \,_1 $
and for $n \,\ge\, 2$ let $\Kn$ be the orthogonal complement of $\ubH
\,_{n-1}$ in $\Hn$.  Then $\Hn \ =\  \ubH \,_{n-1} \, \oplus \, \Kn$
and $\underline{\bf K} \,_m \,\ppd\, \Kn$ if $m \neq n$.  Then \eq{eq51c}
implies that for any ${\bf h} \, \in \, \ubH$ there is a unique sequence
of vectors ${\bf k}_1 ,\ {\bf k}_2 , \  ...$ with ${\bf k}_n \,\in\,
\Kn$ and such that

\begin{equation}
\lim_{{N} \to \infty} \   \|  \, {\bf h} \ -\  \sum_{n=1}^N \ 
{\bf k}_n  \|  \ =\  0 \,.
\label{eq52}\end{equation}
The self-adjointness of $F$ implies that $F \Kn \ \subseteq\ \Kn $
for all $n$, and thus $F|\,\Kn $ is a self-adjoint operator on the
finite-dimensional space $\Kn$.  Therefore $\Kn$ has an orthonormal
basis consisting of eigenvectors of $F|\,\Kn$ (Halmos, 1958, p156
\cite{halmos58}).  Collecting all these eigenvectors for all the $\Kn$
gives an orthonormal set of eigenvectors of $F$ in $\ubH$, and by
\eq{eq52} they constitute an orthonormal basis for $\underline{\bf H}$.

The direct application of the construction \eq{eq51} to the Poincar\'e
problem formulated in section~\ref{sect3} would be to take $\ubH \ =\   \ubLam$
and $F\ =\  L|\,  \ubLam$.  It turns out to be easier to take $\ubH \
=\   \ubPi$ and $F\,=\  L$.  Suppose that $ \ubPi$ contains a sequence
of subspaces $ \ubPi \,_1 ,\   \ubPi \,_2 ,\  ...$ such that \eq{eq51}
is true with $\ubH \ =\   \ubPi$, $\Hn \ =\   \ubPi \,_n$, and $F\ =\  L$.
We claim that then $ \ubLam$ has a complete orthonormal basis consisting
of eigenfunctions of $L|\,  \ubLam$.

To see this, note that \eq{eq51} also holds with $\ubH
= \overline{\,L\ubPi}$, $\Hn = L\ubPi\,_n$, and $F\ = L|\,
\overline{\,L\ubPi}$.  Therefore $\overline{\,L\ubPi}$ has an orthonormal
basis consisting of eigenfunctions of $L$.

Let $ \ubLam \,_0$ be the set of all ${\bf w}$ in $ \ubLam$ such that
$L {\bf w} \ =\   {\bf 0}$.  Greenspan (1968, p40) calls these the
geostrophic motions. Any orthonormal basis for $ \ubLam \,_0$ consists
of eigenvectors of $L$.  Therefore we have an orthonormal basis for $
\ubLam$ consisting of eigenvectors of $L$ if we can prove that

\begin{equation}
\ubLam = \ubLam_0 \oplus \overline{L\ubPi} \,.
\label{eq53}\end{equation}
To prove \eq{eq53}, note first that if ${\bf w} \,\in\,  \ubLam \,_0$
then $\langle \, L {\bf w} \,|\, {\bf v} \, \rangle \ =\  0$ for every
${\bf v} \,\in\,  \ubPi$.  Hence $\langle \, {\bf w} \,|\, L {\bf v}
\, \rangle \ =\  0$ for every ${\bf v} \,\in\,  \ubPi$.  Hence ${\bf
w} \ \ppd\, L  \ubPi$, so ${\bf w} \ \ppd\, \overline{\,L  \ubPi}$.
Thus $ \ubLam \,_0 \ \ppd\,\overline{\,L  \ubPi} $.  Next, suppose
${\bf w} \,\in\,  \ubLam$ and ${\bf w} \ \ppd\, \overline{\,L  \ubPi}$.
Since $L^2 \, {\bf w} \,\in\, \overline{\,L  \ubPi}$, therefore $\langle
\, {\bf w} \,|\, L^2 \, {\bf w} \, \rangle \ =\  0$.  But $\langle \,
{\bf w} \,|\, L^2 \, {\bf w} \, \rangle \ =\ \langle \, L {\bf w} \,|\,
L {\bf w} \, \rangle \,$, so $L {\bf w} \ =\   {\bf 0}$ and ${\bf w}
\,\in\,  \ubLam \,_0$.

\subsection{Polynomial subspaces}

To apply the foregoing general remarks to the Poincar\'e problem,
we set $\ubH \ =\   \ubPi $ and $F\ =\ L$ in \eq{eq51}, and we seek
appropriate spaces $ \ubPi \,_n$ to use as the $\Hn$ in \eq{eq51}.
In the axisymmetric ellipsoid, the Poincar\'e modes are all polynomial
velocity fields (Greenspan, 1968, p64).  This suggests that spaces of
such fields might serve as the $ \ubPi \,_n$.  To describe these spaces
requires some notation.  The origin of coordinates is fixed somewhere on
the axis about which the fluid rotates, and ${\bf r}$ is the position
vector relative to this origin.  Let $ \uPi \,[l,\,l]$ be the set
consisting of 0 and all complex homogeneous polynomials of degree $l$
in ${\bf r}$.  If $l < n$, let $\uPi \,[l,\,n]$ be the set consisting
of 0 and all polynomials whose monomial terms have degrees from $l$ to
$n$ inclusive.  Let $\uPi \,[l,\,\infty ]$ be the set consisting of 0 and
all polynomials whose constituent monomials have degree $l$ or greater.
For any pair of integers $(l,\,n)$ with $l \,\le\, n$, including $n =
\infty$, let $ \ubPi \,[l,\,n]$ denote the set of vector fields whose
Cartesian components are members of $\uPi \,[ l,\,n]\,$.

The arguments to follow will compare the dimensions of
various linear spaces, and these dimension counts begin with the
spaces just described.
By the definition of $\uPi \,[l,\,l]$, it is spanned by the monomials
$x^a y^b z^c$ with $a\,+\,b\,+\,c\ =\ l$.  They are 
linearly independent, and their number is easily seen to be
$(l+1)(l+2)/2$, so

\begin{subequations}
\begin{equation}
{\rm dim} \  \uPi \,[l,\,l] \ =\  (l\,+\,1)(l\,+\,2)/2 \,.
\label{eq54a}\end{equation}
Summing \eq{eq54a} from $l=0$ to $l=n$ gives
\begin{equation}
{\rm dim} \  \uPi \,[\,0,\,n] \ =\  (n\,+\,1)(n\,+\,2)(n\,+\,3)/6\,.
\label{eq54b}\end{equation}
Then ${\rm dim} \  \uPi \,[l,\,n]$ for $l\,\ge\,1$ can be computed from
\begin{equation}
{\rm dim} \  \uPi \,[l,\,n] \ =\  {\rm dim} \  \uPi \,[\,0,\,n] \ -\  {\rm dim} \ 
\uPi \,[\,0,\,l\,-\,1] \,.
\label{eq54c}\end{equation}
\par\noindent
The foregoing formulas hold with $\uPi$ replaced by 
$ \ubPi$ if the right sides of (\ref{eq54}a,b) are multiplied by 3.
In particular
\begin{equation}
{\rm dim} \   \ubPi \,[\,0,\,n] \ =\  (n\,+\,1)(n\,+\,2)(n\,+\,3)/2 \,.
\label{eq54d}\end{equation}
\label{eq54}
\end{subequations}

For later convenience we ignore $  \ubPi \,[\,0,\,0]$ and 
$  \ubPi \,[\,0,\,1]$.  In proving \eq{eq51} we take $ \ubH\ =\ 
 \ubPi $, $F\ =\ L$, and $ \Hn \ =\   \ubPi \,[\,0,\,n+1]$ with 
$n\ =\ 1,\,2, ... $.
Both \eq{eq51a} and \eq{eq51b} are obvious, and \eq{eq51c} is well known
(Korevaar, 1968, p375 \cite{korevaar68}; Courant and Hilbert, 1953,
p68 \cite{CH53}).

It remains only to verify \eq{eq51d} when $F\ =\ L$ and 
$\Hn \ =\   \ubPi \,[\,0,\,n+1]\,$.  We must show that 
if $n\ \ge\ 2$
\begin{equation}
L  \ubPi \,[\,0,\,n] \ \subseteq\   \ubPi \,[\,0,\,n] \,.
\label{eq55}\end{equation}
From \eq{eq314}, clearly
\begin{equation}
R  \ubPi \,[\,0,\,n]\ \subseteq\   \ubPi \,[\,0,\,n]
\label{eq56}\end{equation}
so \eq{eq55} will follow from \eq{eq316b} if it can be shown that

\begin{subequations}
\begin{equation}
\Lambda  \ubPi \,[\,0,\,n] \ \subseteq\   \ubPi \,[\,0,\,n]\,.
\label{eq57a}\end{equation}
Since $\Gamma \,+\, \Lambda \ =\  {\rm I} \,_{  \ubPi } $, 
\eq{eq57a} is equivalent to
\begin{equation}
\Gamma \,  \ubPi \,[\,0,\,n] \ \subseteq\   \ubPi \,[\,0,\,n] \,.
\label{eq57b}\end{equation}
\label{eq57}
\end{subequations}
Thus everything hinges on proving \eq{eq57b}.  Lebovitz (1989) proves
\eq{eq57} directly by constructing explicit polynomial bases for
$ \Lambda  \ubPi \,[\,0,\,n]$ and $ \Gamma  \ubPi \,[\,0,\,n]$
and showing that their total number is $ {\rm dim} \  \ubPi
\,[\,0,\,n]$.  We give here an alternate proof which avoids some 
computation.

\subsection{The case of the ellipsoid}

We now show that \eq{eq57b} is true whenever $E$ is an ellipsoid, axisymmetric
or not.  We take the ellipsoid's principal axes as the coordinate axes,
so that the equation of $ \pd E$ is

\begin{equation}
Ax^2 \ +\  By^2 \ +\  Cz^2 \ =\  1
\label{eq58}\end{equation}
for some positive constants $A,\ B,\ C$.  Then the outward unit normal
to $ \pd E$ is ${\hat {\bf n}} \ =\  {\bf K} / \|{\bf K}\|$ where,
in an obvious notation,

\begin{subequations}
\begin{equation}
{\bf K} \ =\  Ax {\hat {\bf x}} \ +\  By {\hat {\bf y}} \ +\ 
Cz {\hat {\bf z}}
\label{eq59a}\end{equation}
and
\begin{equation}
\| {\bf K} \| \ =\  (A^2 x^2 \,+\, B^2 y^2 \,+\, 
C^2 z^2 )^{{1\over 2}} \,.
\label{eq59b}\end{equation}
\label{eq59}
\end{subequations}

Let $D\ =\  {\bf K} \,\cdot\,  \nabla$, so that
\begin{equation}
D \ =\  Ax\, \pd_x \ +\  By\, \pd_y \ +\  Cz\, \pd_z \,.
\label{eq510}\end{equation}

To prove \eq{eq57b} we choose any ${\bf v} \,\in\,  \ubPi 
\,[\,0,\,n]$ and try 
to show that $\Gamma {\bf v} \,\in\,  \ubPi \,[\,0,\,n]$
when $n\ \ge\ 2$.
We know that $\Gamma {\bf v} \ =\   \nabla \phi$ where $\phi$
solves \eq{eq313}. That is,

\begin{subequations}
\begin{equation}
\nabla^2 \phi \ =\   \nabla \,\cdot\, {\bf v} \ \ \ \ {\rm in}
\ \ \ \ E
\label{eq511a}\end{equation}
\begin{equation}
D \phi \ =\  {\bf K} \,\cdot\, {\bf v} \ \ \ \ {\rm on} \ \ \ \  \pd E \,.
\label{eq511b}\end{equation}
\label{eq511}
\end{subequations}
If we can show that \eq{eq511} has a solution $\phi$ in $\uPi \,[1,\,n\,+\,1]$,
then $ \nabla \phi \,\in\,  \ubPi \,[\,0,\,n]$, and \eq{eq57b}
is established.

An idea of Cartan (1922, p358)\nocite{Cartan22} finds $\phi$.
We note first that if $ {\bf v} \,\in\, {\bf \uPi} \,[\,0,\,n]$ 
then ${\bf K} \,\cdot\, {\bf v} \,\in\, \uPi \,[1,\,n\,+\,1]$. 
Next we claim that
$D:\, \uPi \,[1,\,n+1]$$\,\to\, \uPi \,[1,\,n\,+\,1]$ has an inverse,
$D^{-1} \,:\, \uPi \,[1,\,n\,+\,1] \,\to\, \uPi \,[1,\,n\,+\,1]$.
To see this, observe that the monomials $x^a y^b z^c$
with $1\, \le \ a\,+\,b\,+\,c\, \le \ n\,+\,1$ are a basis for
$ \uPi \,[1,\,n\,+\,1]$ and that
\begin{equation}
D\,x^a y^b z^c \ =\  (Aa \,+\, Bb \,+\, Cc)\,x^a y^b
z^c \,.
\label{eq512}\end{equation}
Since $aA + bB + cC $ is positive, we can divide by it and solve
\eq{eq512} for $D^{-1} \,x^a y^b z^c$.

Now let $\psi \,\in\, \uPi \,[1,\,n-1]$ and consider
the function $\phi$ defined by
\begin{equation}
\phi \ =\  D^{-1} [\, {\bf K} \,\cdot {\bf v} \,+\, (Ax^2 \,+\,
By^2 \,+\, Cz^2 \,-\, 1)\, \psi ] \,.
\label{eq513}\end{equation}
Clearly 
$\phi \,\in\, \uPi \,[1,\,n\,+\,1]$, and $\phi$ satisfies \eq{eq511b}.
Can $\psi$ be chosen in $\uPi \,[1,\,n-1]$ so that $\phi$ also
satisfies \eq{eq511a}?  If so, we have proved \eq{eq57b}.  Thus the question
is whether, given $ {\bf v} \,\in\, {\bf \uPi} \,[\,0,\,n]$,
we can find a $\psi$ in $\uPi \,[1,\,n-1]$ such that
\begin{subequations}
\begin{equation}
T \psi \ =\  \alpha
\label{eq514a}\end{equation}
where
\begin{equation}
T \psi \ =\  \nabla^2 D^{-1} [(Ax^2 \,+\, By^2 \,+\, 
Cz^2 \,-\, 1)\, \psi ]
\label{eq514b}\end{equation}
and
\begin{equation}
\alpha \ =\   \nabla \,\cdot\, {\bf v} \ -\  \nabla^2 D^{-1}
( {\bf K} \,\cdot\, {\bf v} ) \,.
\label{eq514c}\end{equation}
\label{eq514}
\end{subequations}

Define $ G_{n-1} $ to be
the set of all scalar fields $ \alpha $ on $E$ such that 

\begin{subequations}
\begin{equation}
\alpha \ \in\  \uPi \,[\,0,\,n-1]
\label{eq515a}\end{equation}
and
\begin{equation}
\int_E \,dV\, \alpha \ =\  0 \,.
\label{eq515b}\end{equation}
\label{eq515}
\end{subequations}
For any vector field $ {\bf v} $ Gauss's theorem implies \eq{eq515b}
for the $\alpha$ computed from \eq{eq514c}.  If also $ {\bf v} \,\in\,
\ubPi \,[0,\,n]$ then clearly $\alpha$ also satisfies \eq{eq515a}, so
$\alpha\,\in\,G_{{n-1}}\,$.  Therefore, to show that \eq{eq514a} has a
solution $\psi\,\in\, \uPi \,[1,\,n-1]$ it suffices to show that

\begin{equation}
T \uPi \,[1,\,n-1] \ =\  G_{n-1} \,.
\label{eq516}\end{equation}

We establish \eq{eq516} in two stages.  First we prove that 

\begin{subequations}
\begin{equation}
T \uPi \,[1,\,n-1] \ \subseteq\  G_{n-1}
\label{eq517a}\end{equation}
and then we prove that
\begin{equation}
{\rm dim} \  T \uPi \,[1,\, n-1] \ =\  {\rm dim} \  G_{n-1} \,.
\label{eq517b}\end{equation}
\label{eq517}
\end{subequations}
To prove \eq{eq517a}, note that if $\psi \,\in\, \uPi \,[1,\,n-1]\,$
and $ \alpha \ =\  T \psi $ then the definition of $T$, \eq{eq514b}, makes 
\eq{eq515a} obvious, while \eq{eq515b} follows from Gauss's theorem.  To
prove \eq{eq517b}, we note that 

\[ {\rm dim}\, G_{n-1} = n(n+1)(n+2)/6 - 1 = {\rm dim}\,\uPi\,[1,\,n-1]\, ,\]

\noindent so it suffices to prove that $T$ is injective, since in that
case ${\rm dim}\,\uPi\,[1,\,n-1]={\rm dim}\,T\uPi\,[1,\,n-1]$. Thus we
need to show that
$T\,\psi\ =\ 0$ implies $\psi\ =\ 0$.  Let $\phi \ =\  D^{-1} [(Ax^2 \,+\,
By^2 \,+\, Cz^2 \,-\,1)\, \psi ]$.  Then $T \psi \ =\ 0$ implies $\nabla^2
\phi \ =\  0$ everywhere, while obviously $D \phi \ =\  0$ on $ \pd E\,$,
so ${\hat {\bf n}} \,\cdot \,  \nabla \phi \ =\  0$
on $ \pd E$.  Thus $\phi$ is constant in $E$.  Then $(Ax^2 \,+\, By^2
\,+\, Cz^2 \,-\, 1)\, \psi \ =\  D \phi \ =\  0$ in $E$.  Hence $\psi \
=\ 0$ everywhere.

At this point the chain of argument is complete.  We have proved
\eq{eq516} and hence \eq{eq57} when $ \pd E$ is the ellipsoid \eq{eq58},
oriented in any way relative to $ {\bf\Omega}$.  In consequence we have
\eq{eq55}, so that \eq{eq51} is verified when $\ubH \ =\   \ubPi$,
$\Hn \ =\   \ubPi \,[\,0,\,n\,+\,1]$ and $F\ =\ L$.  It follows that
when $ \pd E$ is an ellipsoid then $ \ubLam$ has an orthonormal basis
consisting of velocity fields ${\bf w}_1 ,\  {\bf w}_2 , \  ... $ each
of which is an eigenvector of $L| \,  \ubLam$ and is an inhomogeneous
polynomial in ${\bf r}$.  This last fact makes available Kudlick's
argument (Greenspan, 1968, p61) that $+1$ and $-1$ cannot be eigenvalues
of any ${\bf w}_n$, so all the eigenvalues $\lambda_n$ of $L|\,  \ubLam$
satisfy $-1 < \lambda_n < 1$.

The foregoing demonstration essentially hinges on the fact that the
ellipsoid is a smooth quadratic surface, so that we can work in the
functional spaces of polynomials which are square-integrable and
infinitely differentiable. With a polynomial velocity field of
$\ubPi[0,n]$, we have proved that the projection on the subspace $\ubGa$
is an internal operation, i.e.  $\Gamma(\vv)$ still belongs to
$\ubPi[0,n]$. Since the subspace $\ubLam$ of the mass conservative
velocity field and $\ubGa$ are orthogonal and complementary, it also
means that the projection on $\ubLam$ is also an internal operation for
this polynomial space. However, it is easier to work with vector
velocity fields of $\ubGa$ because these vector fields are irrotational
and simply described by a scalar function. With these remarks the
operator $L$ is also internal in the polynomial space $\ubPi[0,n]$ and
polynomial eigenfunctions are possible.

\section{The Poincar\'e modes}\label{poincaremodes}

\subsection{Known properties}

For an axisymmetric ellipsoid rotating about its axis of symmetry Bryan
(1889) extracted from Poincar\'e (1885) paper a list of particular
polynomial eigenvelocities belonging to the family described in the
preceding section, and expressible in closed form in terms of Legendre
functions.  For the Poincar\'e problem Greenspan (1968) and
\cite{rieutord15}
give a succinct description of such modes.  These Poincar\'e modes are
described in more detail than is usual in the literature in appendix B
of the paper, this in order to count them and to make possible a proof
in the next section that they are complete if supplemented by some
geostrophic modes.

From appendix B, we shall keep in mind that the pressure field
associated with the eigenmodes read:

\begin{equation}
q(s,\phi,z)\ =\ e^{im\phi}\,P_l^m (\sin\xi)
P_l^m (\sin\eta)
\label{B18}\end{equation}
for any given integer $l\geq1$ and $m\in[-l,l]$. In this expression,
$\xi$ and $\eta$ are given as functions of the cylindrical coordinates
$s$ and $z$ by \eq{eq68} and $P_l^m$ are the classical associated Legendre
polynomials. The determination of the eigenfrequency needs
the computation of a root of

\begin{equation}
[\cos\gamma\partial_{\gamma} - mh( \gamma )] P_l^m ( \sin\gamma )  = 0
\label{B17}\end{equation}
with $0 < |\gamma| <  \pi/2$ and where $h(\gamma)$ is given
by \eq{eq66}. Then, the root $\gamma$ serves in the relation between
$\xi$, $\eta$, $s$ and $z$ \eq{eq68} and for the determination of the
eigenfrequency through \eq{eq66}.

For $m\ =\ 0$ the polynomial solutions given by \eq{B18} have an
important peculiarity.  In that case, if $\gamma_0$ solves \eq{B17}
so does $-\gamma_0$, and the two coordinate systems \eq{eq68} generated
from $ \gamma  = \gamma_0 $ and $ \gamma = -\gamma_0$ give the same pressure
function $q$ via \eq{B18}.  However, they give different eigenvalues
$ \lambda $ in \eq{eq66}, equal except for opposite signs.  Hence they
generate different velocity fields $ {\bf v} $ in \eq{vvp}.  In ordinary
eigenvalue problems, the eigenfunction has a unique eigenvalue, so it
is better bookkeeping to regard the velocity field $ {\bf v}
$ rather than the pressure field $q$ as the eigenfunction belonging to
the eigenvalue $\lambda$.

As noted by Cartan (1922), Kudlick (1966) and Greenspan (1968, p65),
the pressure functions \eq{B18} are inhomogeneous polynomials of
degree $l$ in the Cartesian coordinates $x,\,y,\,z$, a fact which can
be verified from \eq{eq615b}.  Hence the velocity field $ {\bf v} $
calculated via \eq{vvp} from the $q$ of \eq{B18} and the $\lambda$ of
\eq{eq66} has Cartesian components which are inhomogeneous polynomials
of degree $l\,-\,1$ in $x,\,y,\,z$.

One other observation will simplify the bookkeeping:  when $m\  \neq \ 
0$, $\gamma = 0$ cannot be a root of \eq{B17} because the left side 
of \eq{B17} is the sum of two terms, one even and one odd in $ \gamma $.
The odd term must vanish at $ \gamma \ =\ 0$, so the even term cannot.
Otherwise $P_l^m ( \mu )$ would have a double zero at $ \mu 
\ =\ 0$.  Being a nonzero solution of a second order linear ordinary
differential equation, $P_l^m $ can have no double zeros.

When $m = 0$, the foregoing argument also shows that $\gamma=0$
cannot be a root of \eq{B17} if $l$ is odd.  If $l$ is even and $m = 0$,
then $\gamma = 0$ must be a root of \eq{B17}.  This produces $ \gamma =
0$, and thus $\lambda = 0$ in \eq{eq66}.  But $\gamma = 0$ cannot
be used in \eq{eq68} to generate a curvilinear coordinate system, so there
is no pressure field \eq{B18} or velocity field \eq{vvp} corresponding
to the root $\eta =0$ of \eq{B17} when $m = 0$ and $l$ is even.
This gap is easily repaired.  For $\lambda = 0$ the pressure field

\begin{subequations}
\begin{equation}
q\ =\ s^l \ =\ (x^2 \ +\ y^2 )^{l/2}
\label{eq619a}\end{equation}
and the velocity field obtained from it via \eq{inert}, not \eq{vvp}, 
\begin{equation}
{\bf v} \ =\ ( \partial_s q)\, {\hat {\bf \phi}} \ =\ l\,s^{l-2} \,(y {\hat {\bf x}} \ -\  x {\hat {\bf y}} )
\label{eq619b}\end{equation}
\label{eq619}
\end{subequations}
are solutions of \eq{bcmass2} and \eq{inert}.  {These are the
classical geostrophic solutions}.
When $l$ is even, \eq{eq619a} is a polynomial
in $x,\,y,\,z$ of degree $l$, and the Cartesian components of \eq{eq619b}
are polynomials of degree $(l\,-\,1)$.  It seems reasonable to assign the
eigenvalue $ \lambda \ =\ 0$ and the pressure and velocity eigenfunctions
\eq{eq619} to the root $\gamma = 0$ of \eq{B17} when $m\ =\ 0$
and $l$ is even.

These bookkeeping conventions permit a simple enumeration of the 
Poincar\'e polynomial solutions of \eq{bcmass2} and \eq{inert}.
For each integer $l\,\ge\,1$ and each integer $m$ in
$-l \,\le\,m\,\le\,l$, let $\eta$ be a root of
\begin{subequations}
\begin{equation}
[\, \cos\eta\,\partial_{\eta} \ -\  mh( \eta ) \,]\ P_l^m 
( \sin\eta ) \ =\  0
\label{eq620a}\end{equation}
\begin{equation}
- \pi /2 <  \eta <  \pi /2 \,.
\label{eq620b}\end{equation}
\label{eq620}
\end{subequations}
Set $\gamma = \eta$ and find $\lambda$ from \eq{eq66}.  Find $q$ and
${\bf v}$ from \eq{eq68}, \eq{B18} and \eq{vvp} except when $\eta = 0$.
The root $\eta = 0$ can appear only when $m = 0$ and $l$ is even.
In that case, find $q$ and ${\bf v}$ from \eq{eq619}.  Any $q$ and ${\bf
v}$ obtained in one of these ways will be called an $(l,m)$-Poincar\'e
pressure polynomial and an $(l,m)$-Poincar\'e velocity polynomial.
An $(l,m)$ pressure polynomial is an inhomogeneous polynomial of
degree $l$ in $x,\,y,\,z$, and the Cartesian components of an $(l,m)$
velocity polynomial are inhomogeneous polynomials of degree $l-1\,$
in $x,\,y,\,z\,$.

The foregoing discussion summarizes very briefly the classical literature
on the Poincar\'e polynomial solutions of \eq{bcmass2}, \eq{inert} when
$\partial E$ is an ellipsoid symmetric about the axis of rotation of
the fluid.  We propose to supplement this classical work with a proof
in section~\ref{cpvp} that the Poincar\'e velocity polynomials are complete.
That proof requires that we have a lower bound for the number $N(l,\,m)$
of $(l,m)$-Poincar\'e velocity polynomials.  Our bookkeeping conventions
assure that $N(l,\,m)$ is just the number of roots of \eq{eq620}.

\subsection{A lower bound for the number of $(l,m)$-Poincar\'e velocity
polytnomials}

To calculate this number, define
$\mu =\sin\eta$ and $g(\mu)=h(\eta)$, so that from \eq{eq66b}

\begin{subequations}
\begin{equation}
g( \mu ) \ =\  [\,1\,-\, \epsilon (1\,-\, \mu^2 )\,]^{{1\over 2}}
\label{eq621a}\end{equation}
where
\begin{equation}
\epsilon \ =\  1 \ -\  (c/a)^2 \,.
\label{eq621b}\end{equation}
measures the flatness of the spheroid.
\label{eq621}
\end{subequations}
Then \eq{eq620} becomes
\begin{subequations}
\begin{equation}
[\,(1\,-\, \mu^2 ) \partial_{\mu} \,-\, mg( \mu )\,]\  
P_l^m ( \mu ) \ =\  0
\label{eq622a}\end{equation}
with
\begin{equation}
-1 <  \mu < 1 \,.
\label{eq622b}\end{equation}
\label{eq622}
\end{subequations}
First, suppose $m\,=\,0$.
Then $l\,+\,1$ applications of Rolle's theorem in the expression of
associated Legendre polynomials, namely, 

\begin{equation}
P_l^m ( \mu )\ =\ (2^l \, l!)^{-1} \,(1- \mu^2 )^{m/2} 
\, \partial_{\mu}^{l+m} \,( \mu^2 \ -\ 1)^l
\label{B15b}\end{equation}
show that
\begin{equation}
N(l,\,0) \ =\  l \,-\,1 \,.
\label{eq623}\end{equation}

Next, suppose $m \neq 0$.  If $\mu$ is a root of \eq{eq622} for this $m$,
then $-\,\mu$ is a root for $-\,m$. As Greenspan (1968, p64) observes,
this means that the Poincar\'e modes with $m \,\neq\, 0$ are traveling
waves. Therefore

\begin{equation}
N(l,\,m) \ =\  N(l,\,-m)
\label{eq624}\end{equation}
and we need calculate $N(l,\,m)$ only when $m>0$.
To this end, define

\begin{equation}
F( \mu ) \ =\  \int_0^{\mu} \, d \zeta \ g ( \zeta )\,(1\,-\, \zeta^2 )^{{-1}} \,,
\label{eq625}\end{equation}
so that \eq{eq622a} becomes

\begin{equation}
\partial_{\mu} \, [\,e^{{-mF(} \mu )} P_l^m ( \mu )\,] \ =\  0 \,.
\label{eq626}\end{equation}
Note that
\[
\frac{g(\zeta)}{1-\zeta^2} = {1\over 2} \,(1\,-\,\zeta )^{-1}
\ +\  {1\over 2} (1\,+\, \zeta )^{-1} \ -\  
\epsilon \,(1\,+\,g( \zeta )\,)^{-1} \,.
\]
so that
$$
F( \mu ) \ =\  {1\over 2} \,\ln (1\,+\, \mu ) \ -\  {1\over 2} \, \ln (1\,-\, \mu ) \ -\  
\ln \, G( \mu )
$$
where
$$
G( \mu ) \ =\  \epsilon \  \int_0^{\mu} \  d \zeta \, [ 1 \,+\, 
g( \zeta )\,]^{-1} \,.
$$
Using \eq{B15b}, we can now write \eq{eq622a} as

\begin{equation}
\partial_{\mu} \,[\,G( \mu )^m (1\,-\,\mu )^m \, \partial_{\mu}^{{l+m}} ( \mu^2 \,-\, 1)^l ] \ =\  0 \,.
\label{eq627}\end{equation}
Applying Rolle's theorem $l+m$ times shows that the $(l-m)$'th degree
polynomial $\partial_{\mu}^{{l+m}} ( \mu^2 \,-\, 1)^l$ has exactly
$l\,-\,m$ simple zeros in $-1 < \mu < 1$.  Therefore the $l$'th degree
polynomial $(1\,-\,\mu )^m \, \partial_{\mu}^{{l+m}} ( \mu^2 \,-\,1)^l$
has only these zeros and $m$ zeros at $\mu \,=\, 1$.  Thus the same is
true of the function $G( \mu )^m (1\,-\, \mu )^m \, \partial_{\mu}^{{l+m}}
( \mu^2 \,-\, 1)^l$.  Then Rolle's theorem gives \eq{eq627} at least $l\,-\,m$
roots in $-1 < \mu < 1$.  Thus

\begin{equation}
N(l,\,m) \ \ge\  l \,-\, |\,m| \ \ \ \ {\rm if} \ \ \ \ m \neq  0 \,.
\label{eq628}\end{equation}
This inequality will suffice in section 7 to prove the completeness
of the Poincar\'e velocity polynomials when $\partial E$ is an
ellipsoid symmetric about the axis of rotation of the fluid.
That proof will produce, as a byproduct, the conclusion that
equality must hold in \eq{eq628}, so

\begin{equation}
N(l,\,m) \ =\  l \,-\, |\,m| \ \ \ \  {\rm if} \ \ \ \ m \neq 0 \,.
\label{eq629}\end{equation}

One interesting consequence of \eq{eq622} is that the eigenvalues
$\lambda$ of the Poincar\'e problem \eq{bcmass2}, \eq{inert} in an
axisymmetric ellipsoid are dense in the interval $-1 < \lambda < 1$.
Indeed, the eigenvalues belonging to $m\,=\,0$ are already dense.  To see
this, observe that for $m\,=\,0$ \eq{eq622a} becomes $\partial_{\mu} P_l^0
( \mu ) \ =\  0$.  An integration by parts and an appeal to Legendre's
equation show that

\beqan
\int_{-1}^1  \,d \mu \ (1\,-\, \mu^2 )\,
\partial_{\mu} P_l^0 ( \mu )\, \partial_{\mu} P_{{l}^\prime}^0 ( \mu )
\nonumber \\
 = l(l\,+\,1)\,  \int_{-1}^1 \,
d \mu \ P_l^0 ( \mu ) \,P_{{l}^\prime}^0 ( \mu )
\eeqan{eq630}
so that the polynomials $\partial_{\mu} P_l^0 ( \mu )$ with $l\ =\
1,\,2,\,3,\,...$ are orthogonal on $-1 < \mu < 1$ with weight function
$(1\,-\, \mu^2 )$.  It follows (Szeg\"o, 1967, p 111) that their zeros
are dense in that interval. \nocite{szego67}

\def\eps{ \epsilon }
\def\buu{ {\bf u} }
\def\bdel{  \nabla }
\def\bq{ \underline{\bf q} }
\def\bQ{ \underline{\bf Q} }
\def\bLAM{  \ubLam }
\def\bGAM{  \ubGa }
\def\Kn{ \underline{\bf K} \,_n }
\def\Hn{ \ubH \,_n }
\def\pd{ \partial }
\def\pdt{ \partial_t }
\def\pdx{ \partial_x }
\def\pdy{ \partial_y }
\def\pdz{ \partial_z }
\def\nomem{ \ \it\hbox{\z\in\h'.15m'\v'.07m'\(sl\v'-.07m'\h'-.15m'} }

\section{Completeness of the Poincar\'e velocity polynomials in an
axisymmetric ellipsoid}\label{cpvp}

The present section proves the claim made in its title: we wish to verify
that the polynomials that have been found by Bryan \cite{bryan1889}
for the spheroid form indeed the complete base that we expect for the
ellipsoid.

\subsection{Dimension of the polynomial subspace $\Lambda\ubPi
\,[\,0,\,n]$}

The proof depends on an appeal to section~\ref{triaxal}.
As noted in that section, $\ubPi$ is the closure of
$ \cup_{n=1}^{\infty}  \ubPi \,[\,0,\,n]$.
Since $\bLAM \ =\  \Lambda  \ubPi $ and $\Lambda$ is
continuous, it follows that

\begin{equation}
\bLAM \ =\  \overline{\, \cup_{n=1}^{\infty} \Lambda  \ubPi \,[\,0,\,n]} \,.
\label{eq71}\end{equation}
Therefore, to prove the completeness of the Poincar\'e velocity
polynomials it sufficies to prove that for each $n$ the Poincar\'e
polynomials of degree $\le\,n$ constitute a basis for $\Lambda  \ubPi
\,[\,0,\,n]$.  In fact, we shall see that they almost constitute an
orthogonal basis.

The first step in the proof is to show that, whatever the 
shape of the fluid volume $E$,

\begin{equation}
{\rm dim} \  \Lambda  \ubPi \,[\,0,\,n] \ =\  n(n+1)(2n+7)/6 \,.
\label{eq72}\end{equation}
Second, when $E$ is an axisymmetric ellipsoid rotating about its axis of
symmetry, of course all the Poincar\'e eigenvelocity fields with degrees
$\le\,n$ are members of $\Lambda  \ubPi \,[\,0,\,n]$, so we finish the
proof by showing that the number of linearly independent Poincar\'e
modes of degree $\le\,n$ is at least \eq{eq72}.

Lebovitz (1989) establishes \eq{eq72} for ellipsoids $E$ by constructing
a particular non-orthonormal polynomial basis for $\Lambda  \ubPi
\,[\,0,\,n]$.  He says (p231, section 7) that such polynomial bases
are available for all shapes $E$. We have not been able to verify this.
Nevertheless, \eq{eq72} {\em is} true for all shapes $E$.  What fails for
some non-ellipsoids (for example, the cube) is \eq{eq57a}.  This does not
rule out the existence of a complete polynomial basis for the Poincar\'e
problem because \eq{eq51} is not an equivalence.

We begin the proof of \eq{eq72} by recalling (Halmos, 1958, p90
\cite{halmos58}) that
if $\underline{\bf Q} $ is any finite dimensional subspace of $ \ubPi $
and $F:\, \ubPi \,\to\, \ubPi $ is linear, and $ {\rm ker\,F|\,\ubQ}$
is the set of all $ {\bf v} \ \in\  \ubQ $
such that $F {\bf v} \ =\   {\bf 0} $, then

\begin{equation}
{\rm dim} \, {\rm ker} \,F|\, \ubQ \ +\  {\rm dim} \, F
\ubQ \ =\  {\rm dim} \, \underline{\bf Q} \,.
\label{eq73}\end{equation}
Next, since $\Lambda$ and $\Gamma$ are orthogonal projectors with
$\Lambda\ +\ \Gamma\ =\ {\rm I}_{{} \ubPi } $, it follows from the 
definitions that 

\begin{equation}
{\rm ker} \, \Lambda\,|\, \ubQ \ =\  \underline{\bf Q} \, \cap \,\Gamma\,
\ubQ \,.
\label{eq74}\end{equation}
Taking $F = \Lambda$ in \eq{eq73} and $\ubQ = \ubPi [0,\,n]$ in \eq{eq73}
and \eq{eq74} gives

\beqan
{\rm dim}(\ubPi [0,n] \cap \Gamma \ubPi [0,n])
+ {\rm dim} \Lambda  \ubPi [0,n] \nonumber \\ 
= {\rm dim} \ubPi \,[\,0,\,n] \,.
\eeqan{eq75}
Then, because of (\ref{eq54}c,d) and \eq{eq75}, we can establish \eq{eq72} by 
showing that 

\begin{equation}
{\rm dim} \ (\,  \ubPi \,[\,0,\,n] \ \cap\  \Gamma \,  \ubPi 
\,[\,0,\,n]\,) \ =\ 
{\rm dim} \  \Pi \,[1,\,n+1] \,.
\label{eq76}\end{equation}
To prove \eq{eq76}, we note first that if $\phi \ \in\  \Pi \,[1,\,n+1]$
and $  \nabla \, \phi \ =\   0 $ then $ \phi \ =\  0\,$.  Thus $  \nabla \
:\  \Pi \,[1,\,n+1] \ \to\   \ubPi \,[\,0,\,n]$ is an injection, so

\begin{equation}
{\rm dim} \  \Pi \,[1,\,n+1] \ =\  {\rm dim} \   \nabla \, \Pi \,[1,\,n+1]
\label{eq77}\end{equation}
Therefore to prove \eq{eq76} it suffices to prove that 

\begin{equation}
 \ubPi \,[\,0,\,n] \ \cap\  \Gamma \,  \ubPi \,[\,0,\,n] \ =\  \bdel \,
\Pi \,[1,\,n+1] \,.
\label{eq78}\end{equation}
The $\supseteq$ half of \eq{eq78} is easy.  If $\phi \,\in\, \Pi \,[1,\,n+1]$,
then $ \nabla \phi \,\in\,  \ubPi \,[\,0,\,n]$, and $\Gamma \,  \nabla
\phi \ =\   \nabla \phi$, so $ \nabla \phi \,\in\, \Gamma \,  \ubPi
\,[\,0,\,n]$.  To prove the $\subseteq$ half of \eq{eq78}, suppose that ${\bf
v} \,\in\,  \ubPi \,[\,0,\,n] \,\cap\, \Gamma \,  \ubPi \,[\,0,\,n]$.
Then $ {\bf v} \ =\  \Gamma \, {\bf v} $, so $ {\bf v} \ =\  \bdel \,
\phi $ for some scalar field $\phi$.  We can calculate $ \phi \,(\,
{\bf r} \,) $ as the line integral of $ {\bf v} $ along a polygonal
curve starting at $  {\bf 0} $, ending at $ {\bf r} $, and consisting of
straight line segments parallel to the coordinate axes.  This calculation
succeeds even if $E$ consists of several disconnected pieces, because
a polynomial known in any open set is uniquely determined in all space,
so the path of integration need not remain in $E$.  Then $ \phi \ \in\
\Pi\,[1,\,n+1]$, and $ {\bf v} \ =\  \bdel \, \phi \ \in\ \bdel \,
\Pi \,[1,\,n+1]$.

\subsection{Number and orthogonality of Poincar\'e polynomials}

\subsubsection{General idea}

Having established \eq{eq72}, now we must count the Poincar\'e modes.
Suppose $\pd E$ is an ellipsoid symmetric about the axis
of rotation of the fluid.
Choose coordinates as in section~\ref{poincaremodes} and let $N(l,\,m)$
be as defined there.  That is, for any integers $l,\,m$ with $l\,\ge\,1$
and $|m|\,\le\,l, \ \ N(l,\,m)$ is the number of $(l,m)$-Poincar\'e
velocity polynomials, and also the number of roots of \eq{eq620}.
Let $\eta\,_{{l,\,m,} \nu }$ be those roots, with $1\ \le\ \nu\
\le\ N(l,\,m)$.  Let $\lambda\,_{{l,\,m,} \nu }$ be the eigenvalues
obtained by setting $\gamma\ =\ \eta_{{l,\,m,} \nu }$ in \eq{eq66}.
Let $ {\bf v} \,_{{l,m} \nu}$ be the corresponding $(l,m)$-Poincar\'e
velocity polynomials, obtained either from \eq{eq619b} or from \eq{vvp},
\eq{eq68} and \eq{B18}.  Then for all $l,\ m,\ \nu$

\begin{equation}
L {\bf v} \,_{{l,\,m,} \nu} \ =\ \lambda\,_{{l,\,m,} \nu} \, 
{\bf v} \,_{{l,\,m,} \nu}
\label{eq79}\end{equation}
and

\begin{equation}
{\bf v} \,_{{l,\,m,} \nu} \ \in \   \ubPi \,
[\,0,\,l-1] \,.
\label{eq710}\end{equation}
We propose to prove that, after a modest amount of Gram-Schmidt
orthogonalization, the ${\bf v} \,_{{l,\,m,} \nu}$ with
$l\,\le\,n+1$ provide an orthogonal basis for $\Lambda  \ubPi \,[\,0,\,n]$.
We make no attempt to normalize these eigenvelocities by finding
$ \|  \, {\bf v} \,_{{l,\,m,} \nu} \,  \|  $.

The proof requires two steps:
(i) to show that the number of ${\bf v} \,_{{l,\,m,} \nu}$ with
$l \,\le\, n\,+\,1$ is at least
${\rm dim} \  \Lambda  \ubPi \,[\,0,\,n]$;
(ii) to show that the ${\bf v} \,_{{l,\,m,} \nu}$ are
linearly independent.
Step (ii) will be accomplished by showing that most of the
${\bf v} \,_{{l,\,m,} \nu}$ are mutually orthogonal and by 
dealing with the exceptions.

\subsubsection{Poincar\'e polynomials are numerous enough}

Step (i) requires counting the Poincar\'e velocity polynomials
${\bf v} \,_{{l,\,m,} \nu}$ for which $l \,\le\, n\,+\,1$.
Their number is obviously
$$
\sum_{l=1}^{n+1} \  \sum_{m=-l}^l \  N(l,\,m)\,,
$$
and, by \eq{eq623}, \eq{eq624} and \eq{eq628}, we know that 

\[ \sum_{m=-l}^l N(l,\,m) \geq l^2-1\;. \]
If we recall that 

\[ \sum_{l=0}^{n+1} (l+1)(l+2) = (n+1)(n+2)(n+3)/3\]
then it turns out that

\begin{equation}
\sum_{l=1}^{n+1} \  \sum_{m=-l}^l \  N (l,\,m) \ \ge\ 
n(n+1)\,(2n+7)\,/\,6\,.
\label{eq711}
\end{equation}
Comparing \eq{eq711} with \eq{eq72}, we see that step (i) is complete.
If we can carry out step (ii), then the $\ge$ in \eq{eq711} must
be an equality.  Hence the same must be true in \eq{eq628}, which
parenthetically proves \eq{eq629}.

\subsubsection{Orthogonality of Poincar\'e polynomials}

It remains to complete step (ii).  As noted by Greenspan (1965; 1968,
p53) and Kudlick (1966),

\begin{subequations}
\begin{equation}
\langle \  {\bf v} \,_{{l,\,m,} \nu} \,|\, {\bf v} \,_{{l}^\prime , m^\prime , \nu^\prime} \, \rangle \ =\  0
\label{eq712a}\end{equation}
whenever
\begin{equation}
\lambda\,_{{l,\,m,} \nu} \neq  \lambda\,_{{l}^\prime ,\,m^\prime , 
\nu^\prime} \,.
\label{eq712b}\end{equation}
\label{eq712}
\end{subequations}
This fact is also evident from the observation that each $ \lambda
\,_{{l,\,m,} \nu}\,$ and $ {\bf v} \,_{{l,\,m,} \nu} $ constitute an
eigenvalue-eigenvector pair of the self-adjoint operator $L\,:\,  \ubPi
\,\to\,  \ubPi$.  There remains the possibility that $\lambda \,_{{l,\,m,}
\nu} \ =\  \lambda \,_{{l}^\prime ,\,m^\prime , \nu^\prime}$ even
though $(l,\,m,\, \nu ) \neq  ( l^\prime ,\,m^\prime ,\, \nu^\prime )$.

\subsubsection{The case of accidental degeneracy}

The foregoing case is called an accidental degeneracy. The question is
to check that even in that case the two eigenmodes are still orthogonal,
namely \eq{eq712a} is still verified.

To deal with this difficulty, we consider other ways of assuring
\eq{eq712a} besides \eq{eq712b}.  For example, \eq{vvp} and \eq{B18}
assure \eq{eq712a} when $m \,\neq\, m^\prime$.

Finally, suppose that $m\ =\ m^\prime $ and 
$(l,\, \nu ) \,\neq\, (l^\prime ,\, \nu^\prime )$ but

\begin{subequations}
\begin{equation}
\lambda \,_{{l,\,m,} \nu} \ =\  \lambda \,_{{l}^\prime ,\,m, \nu^\prime} \,.
\label{eq713a}\end{equation}
When this happens we must have

\begin{equation}
l \neq  l^\prime
\label{eq713b}\end{equation}
\label{eq713}
\end{subequations}
because if $l = l^\prime$ then \eq{eq713a} implies $\nu = \nu^\prime$.
If we do have \eq{eq713} then $\lambda \,_{{l,\,m,} \nu}$ and
$\lambda_{l^\prime ,\,m, \nu^\prime}$ produce the same $\gamma$ in
\eq{eq65} and the same coordinate system in \eq{eq68}.  Therefore the
roots $\mu$ and $\mu^\prime$ of \eq{eq622} must be the same for $l$ and
$l^\prime$ and the given $m$.  But from \eq{eq621} $g$ is a function
of $\eps$ as well as $\mu$.  Suppose we ask how $\mu ,\  \mu^\prime$
and hence ${\bf v} \,_{{l,\,m,} \nu}$ and ${\bf v} \,_{{l}^\prime ,\,m,
\nu^\prime}$ vary as we change $\eps$ slightly.  From \eq{eq622a},
$\pd_{\eps} \, \mu$ is given by

\beqan
[\pd_\mu (1-\mu^2)\pd_{\mu} P_l^m -
mg_{\eps} \pd_{\mu} P_l^m - mP_l^m \pd_{\mu}
g_{\eps} ] \pd_{\eps} \mu =\qquad\nonumber \\  m\,P_l^m \,
\pd_{\eps} \,g_{\eps} \,.\qquad
\eeqan{eq714}
Here the terms in $g_{\eps}$ can be calculated from \eq{eq621a}, $\pd_{\mu}
(1\,-\, \mu^2 )\, \pd_{\mu} P_l^m$ can be expressed in terms of $P_l^m$
by means of Legendre's equation, and when $\mu$ is a root of \eq{eq622}
then $\pd_{\mu} P_l^m$ can be expressed in terms of $P_l^m$.  These
substitutions convert \eq{eq714} into

\beqan
2P_l^m ( \mu )\,[\,l(l+1)g_{\eps} \,-\, \eps \,m^2 g_{\eps} \,+\, \eps
\,m \mu \,]\  \pd_{\eps} \, \mu \ =\qquad\nonumber \\
m (1-\mu^2) \,P_l^m (\mu) \,.\qquad
\eeqan{eq715}
Expression \eq{B15b} of Legendre polynomials and the argument before
equation \eq{eq628} establish that $P_l^m ( \mu )$ has no multiple zeroes.
Therefore, at a root of \eq{eq622} with $m\,\neq\, 0$ we must have
$P_l^m ( \mu ) \,\neq\, 0$.  Hence, when $m \,\neq\, 0$ we can cancel
$P_l^m ( \mu )$ from \eq{eq715} and obtain a formula for $\pd_{\eps} \,
\mu$ in which no terms depend on $l$ except for $l(l+1)$ on the left.
Since $l\,\neq\, l^\prime$, it follows that if $m\,\neq\, 0$ then

\begin{equation}
\pd_{\eps} \, \mu \neq  \pd_{\eps} \, \mu^\prime \,.
\label{eq716}\end{equation}
Therefore, if $m\,\neq\,0$ and $\eps$ is slightly altered, the
eigenvalues of $L$ belonging to ${\bf v} \,_{{l,\,m,} \nu}$ and ${\bf v}
\,_{{l}^\prime ,\,m, \nu^\prime}$ will become different and we will have
\eq{eq712}.  But from \eq{eq68} and \eq{B18}, ${\bf v} \,_{{l,\,m,} \nu}$
and ${\bf v} \,_{{l}^\prime ,\,m, \nu^\prime}$ depend continuously on
$\eps$, so \eq{eq712a} remains true even at the original value of $\eps$
where \eq{eq712b} fails.  From \eq{eq715}, this argument will break down if $m\ =\ 0$,
and that case must now be considered.  All other Poincar\'e velocity
polynomials are orthogonal to each other and to those with $m\ =\ 0$.

When $m\ =\ 0$ there are two kinds of Poincar\'e velocity polynomials
${\bf v} \,_{{l,0,} \nu}$, the proper (non-geostrophic) ones and, for
even $l$, the geostrophic ones.  There are proper Poincar\'e velocity
polynomials with $m\ =\ 0$ only for $l\,\ge\,3$.  By \eq{eq53}, all the proper
ones have nonzero eigenvalues $\lambda$, while all the geostrophic ones
have $\lambda \ =\ 0$.  Therefore, as already noted by Greenspan (1965;
1968, p54) and Kudlick (1966), the proper and geostrophic Poincar\'e
polynomials are orthogonal to one another, and we can consider them
separately.

First consider the proper Poincar\'e velocity polynomials with $m\ =\ 0$.
The $\gamma$'s needed to generate their coordinate systems \eq{eq68}
and pressure fields \eq{B18} are obtained from $\sin\,\gamma\ =\ \mu$,
where $\mu$ is a root of \eq{eq622} with $m\ =\ 0$, i.e.,

\begin{equation}
\pd_{\mu} P_l^0 ( \mu ) \ =\  0 \,.
\label{eq717}\end{equation}
For each fixed $l$, all the different roots of \eq{eq717} generate
different eigenvalues $\lambda$ and hence mutually orthogonal
Poincar\'e velocity polynomials.
The only trouble comes when $l\,\neq\,l^\prime$ and
$\pd_{\mu} P_l^0 ( \mu )$ and
$\pd_{\mu} P_{{l}^\prime}^0 ( \mu )$ have a common zero,
$\mu_0$.
We know no proof that rules this out, but if it does happen then all
the Poincar\'e velocity polynomials produced by the different $l$ which
make $\mu_0$ a root of \eq{eq717} will be orthogonal to all other Poincar\'e
velocity polynomials.  They are linearly independent, being polynomials
of different degrees, so they can always be orthogonalized by the
Gram-Schmidt process.  Perhaps one could prove them mutually orthogonal
by perturbing $\pd E$ into a slightly non-axisymmetric ellipsoid and
using another continuity argument on \eq{eq714}.  But this would require
a discussion of the Lam\'e functions used to produce the analogue of
\eq{B18} in a triaxial ellipsoid (Poincar\'e 1885; Cartan, 1922).

We now consider the geostrophic velocity polynomials \eq{eq619b}.  They are
obviously not mutually orthogonal, but are clearly linearly independent,
being polynomials of different degrees.  This finishes the proof that the
Poincar\'e velocity polynomials are linearly independent, and accomplishes
step (ii) of the overall argument.  Thus the Poincar\'e velocity
polynomials are complete in $\bLAM$ for an axisymmetric ellipsoid $E$.

\subsubsection{Orthogonalized geostrophic velocity polynomials}

Although not necessary for the foregoing argument, it  may be interesting
to note that the Gram-Schmidt orthogonalization of the geostrophic
velocity polynomials can be carried out explicitly.  Write \eq{eq619b} as

\begin{subequations}
\begin{equation}
{\bf v}_l \ =\  C_l \,s\,f_n 
(\,s^2 / a^2 \,)\,  \hat{\mathbf \phi} \,, \ \ \ \ 
l\ =\ 2,\ 4,\ 6, ...
\label{eq718a}\end{equation}
where $C_l$ is a constant, $n\ =\ l/2\,-\,1\,$, and

\begin{equation}
f_n ( \sigma ) \ =\  \sigma^n \,.
\label{eq718b}\end{equation}
\end{subequations}
Then a little calculation gives

\begin{equation}
\langle \ {\bf v}_l \,|\, {\bf v}_{l^\prime} \,
\rangle \ =\  C_{{ll}^\prime} \  \int_0^1 \, d \sigma \,(1- \sigma )^{{1\over 2}} \sigma f_n ( \sigma ) f_{{n}^\prime} ( \sigma )
\label{eq719}\end{equation}
where $n = l/2-1$, $n^\prime = l^\prime/2-1$ and
$ C_{{ll}^\prime}$ is another constant.
Thus Gram-Schmidt orthogonalizing the geostrophic velocity polynomials
${\bf v}_2 \,,\  {\bf v}_4 \,,\  {\bf v}_6 \,,\  ... $
amounts to orthogonalizing the monomials 
$1,\ \sigma ,\ \sigma^2 ,\  ... $ 
on the interval $0\,\le\, \sigma \,\le\, 1$ with the weighting function
$(1- \sigma )^{{1\over 2}} \sigma$.
The resulting orthogonalized polynomials in $\sigma$ are
$P_n^{{(} \alpha , \beta )} (2 \sigma -1)$, where
$n\ =\  l /2-1$, $\alpha \ =\  {1\over 2}$, $\beta \ =\ 1$, and
$P_n^{{(} \alpha , \beta )}$ 
is a Jacobi polynomial (Szeg\"o, 1967, p58).  Thus the orthogonalized
geostrophic velocity polynomials can be taken as

\begin{equation}
\tilde {\bf v}_l \ =\ (l+1)s P_n^{{(} \alpha , \beta )} \,(2s^2 / a^2 -1
)\,  \hat{\bf \phi}
\label{eq720}\end{equation}
where $\alpha \ =\  {1\over 2} \,$, $\beta \ =\  1$ and $n\ =\ l/2\,-\,1\,$.
The corresponding pressure polynomials $\tilde q_l$ are
related to $\tilde {\bf v}_l$ by

\begin{equation}
\tilde{\bf v}_l \ =\  ( \pd_s \tilde q_l )\, \hat{\bf\phi}
\label{eq721}\end{equation}
so (Szeg\"o 1967, p63) we can take

\begin{equation}
\tilde q_l \ =\  a^2 P_n^{(\alpha , \beta)} 
\,(\,2s^2 / a^2 -1) \,.
\label{eq722}\end{equation}
with $\alpha \ =\ - {1\over 2} \,$, $\beta \ =\  1$ and $n\ =\ l/2\,$.

\def\mbar{ {{\   \over m }} }
\def\eps{ \epsilon }
\def\buu{ {\bf u} }
\def\bdel{  \nabla }
\def\bq{ \underline{\bf q} }
\def\bQ{ \ubQ }
\def\bLAM{  \ubLam }
\def\bGAM{  \ubGa }
\def\Kn{ \underline{\bf K} \,_n }
\def\Hn{ \ubH \,_n }
\def\pd{ \partial }
\def\pdt{ \partial_t }
\def\pdx{ \partial_x }
\def\pdy{ \partial_y }
\def\pdz{ \partial_z }
\def\nomem{ \ \it\hbox{\z\in\h'.15m'\v'.07m'\(sl\v'-.07m'\h'-.15m'} }

\section{Conclusions}

In this work we first demonstrated that the Poincar\'e problem, which
governs the inertial oscillations of a rotating fluid, can be formulated
in the space of square-integrable functions without any hypothesis on the
continuity or differentiability of the velocity fields. This formulation
makes available many results of functional analysis. First, while
restricting the velocity field to those that verify incompressibility
and boundary conditions, in other words restricting the velocity fields
to a Hilbert sub-space of the square-integrable vector fields, we could
formulate the Poincar\'e problem as a simple eigenvalue problem namely
$L\vv=\lambda\vv$ showing in passing that the velocity field is the
appropriate variable, rather than the pressure, for this formulation. It
turns out that the operator $L$ is bounded and self-adjoint of norm less
or equal to unity. Hence, the spectrum of $L$ is real and occupies the
interval $[-1,+1]$ of the real axis of the complex frequency plane. A
theorem of functional analysis \cite[e.g.][]{kreyszig78} states that
the residual spectrum of such an operator is empty. Hence, the interval
$[-1,+1]$ is shared by the eigenvalues (the point spectrum) and the
continuous spectrum, the two sets being disjoint and complementary. This
first part gives the general framework that can be used to analyse the
Poincar\'e problem in any type of volumes.

From the foregoing background, we could show that the inertial modes of a
rotating fluid contained in an ellipsoid are polynomial velocity fields
and form a complete base for square-integrable vector fields defined over this
volume. We thus confirm in an independent and more direct way a result
of Lebovitz \cite{lebovitz89}. We also show that the inertial modes of a
spheroid, first obtained by Bryan \cite{bryan1889}, form the expected base
when they are completed by the geostrophic modes. We here confirm,
independently, the same result obtained for the sphere by Ivers et al.
\cite{IJW15}.

Our work shares many results with those obtained in \cite{IJW15}, but
these authors restricted, at the outset, their analysis to continuously
differentiable velocity fields and exhibit the completeness of the
inertial base for the sphere only. In their conclusion they observe that
they could have used an extension of their functional space so as to
use a Hilbert space, and the ensuing results of functional analysis. Our
work thus gives a follow up of this conclusion, but show in addition
that the mere Hilbert space of square-integrable functions is sufficient
for that (instead of the closure of the set of  once continuously
differentiable functions). However, both works shed light on the various
properties of the Poincar\'e problem.

Because Poincar\'e problem is hyperbolic with boundary conditions, thus
ill-posed, the geometry of the container is crucial to the properies of
the eigenspectrum. As shown in \cite[][]{RGV01} information propagated
by characteristics has to be consistent to lead to regular solutions. To
give a physical picture, hyperbolic problem are well-posed with initial
conditions, while here we impose initial and final conditions, which
may not be compatible. Hence, each geometry is a specific case. Except
the ellipsoid and the annular channel \cite{cui_etal14}, it is unknown
whether the Poincar\'e problem has a complete set of eigenvelocities.
Two non-ellipsoidal examples have been considered: the cube and the
spherical shell, but the proof of (in)completeness remained elusive. In
view of the results of Rieutord et al. \cite{RGV01} for the spherical
shell and Nurijanyan et al. \cite{nurijanyan_etal13} for the rectangular
parallelepiped, it may well be that the eigenvalue spectrum is almost
empty for both of these volumes. On the other hand we know since Kelvin
\cite{kelvin1880} that the cylinder admits eigenmodes but the completeness
of their set remains an open question. The present work may give a route
towards the answer.

\bigskip
\noindent {\it Historical note:} The main body of this work was written by GB
in the early 1990's but, when submitted to journals in 1993, did not
meet its readership. Twenty years after, the subject of oscillations of
rotating fluids has been strongly revived and after the demonstration of
\cite{IJW15}, the present work sheds new light on the mathematical
questions associated with inertial modes. The contribution of MR to the
original work has been in updating the introduction and conclusion,
and making the text less difficult when possible.

\begin{acknowledgements}
We are grateful to prof. Stefan Llewellyn Smith for making this
work possible.
We also thank the referee for detailed and patient readings of the
manuscript, which very much helped us in clarifying the presentation of
the heart of the text. NSF grant EAR 89-07988 and NASA grant NAG 5-818
supported parts of this work.
\end{acknowledgements}

\bibliographystyle{apsrev4-1}
\bibliography{../../biblio/bibnew}

\begin{thebibliography}{42}%
\makeatletter
\providecommand \@ifxundefined [1]{%
 \@ifx{#1\undefined}
}%
\providecommand \@ifnum [1]{%
 \ifnum #1\expandafter \@firstoftwo
 \else \expandafter \@secondoftwo
 \fi
}%
\providecommand \@ifx [1]{%
 \ifx #1\expandafter \@firstoftwo
 \else \expandafter \@secondoftwo
 \fi
}%
\providecommand \natexlab [1]{#1}%
\providecommand \enquote  [1]{``#1''}%
\providecommand \bibnamefont  [1]{#1}%
\providecommand \bibfnamefont [1]{#1}%
\providecommand \citenamefont [1]{#1}%
\providecommand \href@noop [0]{\@secondoftwo}%
\providecommand \href [0]{\begingroup \@sanitize@url \@href}%
\providecommand \@href[1]{\@@startlink{#1}\@@href}%
\providecommand \@@href[1]{\endgroup#1\@@endlink}%
\providecommand \@sanitize@url [0]{\catcode `\\12\catcode `\$12\catcode
  `\&12\catcode `\#12\catcode `\^12\catcode `\_12\catcode `\%12\relax}%
\providecommand \@@startlink[1]{}%
\providecommand \@@endlink[0]{}%
\providecommand \url  [0]{\begingroup\@sanitize@url \@url }%
\providecommand \@url [1]{\endgroup\@href {#1}{\urlprefix }}%
\providecommand \urlprefix  [0]{URL }%
\providecommand \Eprint [0]{\href }%
\providecommand \doibase [0]{http://dx.doi.org/}%
\providecommand \selectlanguage [0]{\@gobble}%
\providecommand \bibinfo  [0]{\@secondoftwo}%
\providecommand \bibfield  [0]{\@secondoftwo}%
\providecommand \translation [1]{[#1]}%
\providecommand \BibitemOpen [0]{}%
\providecommand \bibitemStop [0]{}%
\providecommand \bibitemNoStop [0]{.\EOS\space}%
\providecommand \EOS [0]{\spacefactor3000\relax}%
\providecommand \BibitemShut  [1]{\csname bibitem#1\endcsname}%
\let\auto@bib@innerbib\@empty
\bibitem [{\citenamefont {Bryan}(1889)}]{bryan1889}%
  \BibitemOpen
  \bibfield  {author} {\bibinfo {author} {\bibfnamefont {G.}~\bibnamefont
  {Bryan}},\ }\href@noop {} {\bibfield  {journal} {\bibinfo  {journal} {Phil.
  Trans. R. Soc. Lond.}\ }\textbf {\bibinfo {volume} {180}},\ \bibinfo {pages}
  {187} (\bibinfo {year} {1889})}\BibitemShut {NoStop}%
\bibitem [{\citenamefont {Friedlander}\ and\ \citenamefont
  {Siegmann}(1982)}]{FS82a}%
  \BibitemOpen
  \bibfield  {author} {\bibinfo {author} {\bibfnamefont {S.}~\bibnamefont
  {Friedlander}}\ and\ \bibinfo {author} {\bibfnamefont {W.}~\bibnamefont
  {Siegmann}},\ }\href@noop {} {\bibfield  {journal} {\bibinfo  {journal} {J.
  Fluid Mech.}\ }\textbf {\bibinfo {volume} {114}},\ \bibinfo {pages} {123}
  (\bibinfo {year} {1982})}\BibitemShut {NoStop}%
\bibitem [{\citenamefont {Dintrans}\ \emph {et~al.}(1999)\citenamefont
  {Dintrans}, \citenamefont {Rieutord},\ and\ \citenamefont
  {Valdettaro}}]{DRV99}%
  \BibitemOpen
  \bibfield  {author} {\bibinfo {author} {\bibfnamefont {B.}~\bibnamefont
  {Dintrans}}, \bibinfo {author} {\bibfnamefont {M.}~\bibnamefont {Rieutord}},
  \ and\ \bibinfo {author} {\bibfnamefont {L.}~\bibnamefont {Valdettaro}},\
  }\href@noop {} {\bibfield  {journal} {\bibinfo  {journal} {J. Fluid Mech.}\
  }\textbf {\bibinfo {volume} {398}},\ \bibinfo {pages} {271} (\bibinfo {year}
  {1999})}\BibitemShut {NoStop}%
\bibitem [{\citenamefont {{Ogilvie}}\ and\ \citenamefont {{Lin}}(2004)}]{OL04}%
  \BibitemOpen
  \bibfield  {author} {\bibinfo {author} {\bibfnamefont {G.~I.}\ \bibnamefont
  {{Ogilvie}}}\ and\ \bibinfo {author} {\bibfnamefont {D.~N.~C.}\ \bibnamefont
  {{Lin}}},\ }\href@noop {} {\bibfield  {journal} {\bibinfo  {journal} {ApJ}\
  }\textbf {\bibinfo {volume} {610}},\ \bibinfo {pages} {477} (\bibinfo {year}
  {2004})}\BibitemShut {NoStop}%
\bibitem [{\citenamefont {Ogilvie}(2005)}]{O05}%
  \BibitemOpen
  \bibfield  {author} {\bibinfo {author} {\bibfnamefont {G.}~\bibnamefont
  {Ogilvie}},\ }\href@noop {} {\bibfield  {journal} {\bibinfo  {journal} {J.
  Fluid Mech.}\ }\textbf {\bibinfo {volume} {543}},\ \bibinfo {pages} {19}
  (\bibinfo {year} {2005})}\BibitemShut {NoStop}%
\bibitem [{\citenamefont {Rieutord}\ and\ \citenamefont
  {Valdettaro}(2010)}]{RV10}%
  \BibitemOpen
  \bibfield  {author} {\bibinfo {author} {\bibfnamefont {M.}~\bibnamefont
  {Rieutord}}\ and\ \bibinfo {author} {\bibfnamefont {L.}~\bibnamefont
  {Valdettaro}},\ }\href@noop {} {\bibfield  {journal} {\bibinfo  {journal} {J.
  Fluid Mech.}\ }\textbf {\bibinfo {volume} {643}},\ \bibinfo {pages} {363}
  (\bibinfo {year} {2010})}\BibitemShut {NoStop}%
\bibitem [{\citenamefont {{Ogilvie}}(2014)}]{O14}%
  \BibitemOpen
  \bibfield  {author} {\bibinfo {author} {\bibfnamefont {G.~I.}\ \bibnamefont
  {{Ogilvie}}},\ }\href@noop {} {\bibfield  {journal} {\bibinfo  {journal}
  {Ann. Rev. Astron. Astrophys.}\ }\textbf {\bibinfo {volume} {52}},\ \bibinfo
  {pages} {171} (\bibinfo {year} {2014})}\BibitemShut {NoStop}%
\bibitem [{\citenamefont {{Sauret}}\ and\ \citenamefont {{Le
  Diz{\`e}s}}(2013)}]{SLD13}%
  \BibitemOpen
  \bibfield  {author} {\bibinfo {author} {\bibfnamefont {A.}~\bibnamefont
  {{Sauret}}}\ and\ \bibinfo {author} {\bibfnamefont {S.}~\bibnamefont {{Le
  Diz{\`e}s}}},\ }\href@noop {} {\bibfield  {journal} {\bibinfo  {journal} {J.
  Fluid Mech.}\ }\textbf {\bibinfo {volume} {718}},\ \bibinfo {pages} {181}
  (\bibinfo {year} {2013})}\BibitemShut {NoStop}%
\bibitem [{\citenamefont {{Zhang}}\ \emph {et~al.}(2012)\citenamefont
  {{Zhang}}, \citenamefont {{Chan}},\ and\ \citenamefont
  {{Liao}}}]{Zhang_etal12}%
  \BibitemOpen
  \bibfield  {author} {\bibinfo {author} {\bibfnamefont {K.}~\bibnamefont
  {{Zhang}}}, \bibinfo {author} {\bibfnamefont {K.~H.}\ \bibnamefont {{Chan}}},
  \ and\ \bibinfo {author} {\bibfnamefont {X.}~\bibnamefont {{Liao}}},\
  }\href@noop {} {\bibfield  {journal} {\bibinfo  {journal} {Journal of Fluid
  Mechanics}\ }\textbf {\bibinfo {volume} {692}},\ \bibinfo {pages} {420}
  (\bibinfo {year} {2012})}\BibitemShut {NoStop}%
\bibitem [{\citenamefont {{Zhang}}\ \emph {et~al.}(2013)\citenamefont
  {{Zhang}}, \citenamefont {{Chan}}, \citenamefont {{Liao}},\ and\
  \citenamefont {{Aurnou}}}]{Zhang_etal13}%
  \BibitemOpen
  \bibfield  {author} {\bibinfo {author} {\bibfnamefont {K.}~\bibnamefont
  {{Zhang}}}, \bibinfo {author} {\bibfnamefont {K.~H.}\ \bibnamefont {{Chan}}},
  \bibinfo {author} {\bibfnamefont {X.}~\bibnamefont {{Liao}}}, \ and\ \bibinfo
  {author} {\bibfnamefont {J.}~\bibnamefont {{Aurnou}}},\ }\href@noop {}
  {\bibfield  {journal} {\bibinfo  {journal} {Journal of Fluid Mechanics}\
  }\textbf {\bibinfo {volume} {720}},\ \bibinfo {pages} {212} (\bibinfo {year}
  {2013})}\BibitemShut {NoStop}%
\bibitem [{\citenamefont {Greenspan}(1968)}]{Green69}%
  \BibitemOpen
  \bibfield  {author} {\bibinfo {author} {\bibfnamefont {H.~P.}\ \bibnamefont
  {Greenspan}},\ }\href@noop {} {\emph {\bibinfo {title} {{The Theory of
  Rotating Fluids}}}}\ (\bibinfo  {publisher} {Cambridge University Press},\
  \bibinfo {year} {1968})\ p.\ \bibinfo {pages} {327 pp.}\BibitemShut {Stop}%
\bibitem [{\citenamefont {Hadamard}(1932)}]{hadamard32}%
  \BibitemOpen
  \bibfield  {author} {\bibinfo {author} {\bibfnamefont {J.}~\bibnamefont
  {Hadamard}},\ }\href@noop {} {\emph {\bibinfo {title} {{Le probl\`eme de
  Cauchy et les \'equations aux d\'eriv\'ees partielles lin\'eaires
  hyperboliques}}}}\ (\bibinfo  {publisher} {Hermann},\ \bibinfo {year}
  {1932})\BibitemShut {NoStop}%
\bibitem [{\citenamefont {Rieutord}\ \emph {et~al.}(2001)\citenamefont
  {Rieutord}, \citenamefont {Georgeot},\ and\ \citenamefont
  {Valdettaro}}]{RGV01}%
  \BibitemOpen
  \bibfield  {author} {\bibinfo {author} {\bibfnamefont {M.}~\bibnamefont
  {Rieutord}}, \bibinfo {author} {\bibfnamefont {B.}~\bibnamefont {Georgeot}},
  \ and\ \bibinfo {author} {\bibfnamefont {L.}~\bibnamefont {Valdettaro}},\
  }\href@noop {} {\bibfield  {journal} {\bibinfo  {journal} {J. Fluid Mech.}\
  }\textbf {\bibinfo {volume} {435}},\ \bibinfo {pages} {103} (\bibinfo {year}
  {2001})}\BibitemShut {NoStop}%
\bibitem [{\citenamefont {Hollerbach}\ and\ \citenamefont
  {Kerswell}(1995)}]{HK95}%
  \BibitemOpen
  \bibfield  {author} {\bibinfo {author} {\bibfnamefont {R.}~\bibnamefont
  {Hollerbach}}\ and\ \bibinfo {author} {\bibfnamefont {R.}~\bibnamefont
  {Kerswell}},\ }\href@noop {} {\bibfield  {journal} {\bibinfo  {journal} {J.
  Fluid Mech.}\ }\textbf {\bibinfo {volume} {298}},\ \bibinfo {pages} {327}
  (\bibinfo {year} {1995})}\BibitemShut {NoStop}%
\bibitem [{\citenamefont {Rieutord}\ and\ \citenamefont
  {Valdettaro}(1997)}]{RV97}%
  \BibitemOpen
  \bibfield  {author} {\bibinfo {author} {\bibfnamefont {M.}~\bibnamefont
  {Rieutord}}\ and\ \bibinfo {author} {\bibfnamefont {L.}~\bibnamefont
  {Valdettaro}},\ }\href@noop {} {\bibfield  {journal} {\bibinfo  {journal} {J.
  Fluid Mech.}\ }\textbf {\bibinfo {volume} {341}},\ \bibinfo {pages} {77}
  (\bibinfo {year} {1997})}\BibitemShut {NoStop}%
\bibitem [{\citenamefont {Rieutord}\ \emph {et~al.}(2000)\citenamefont
  {Rieutord}, \citenamefont {Georgeot},\ and\ \citenamefont
  {Valdettaro}}]{RGV00}%
  \BibitemOpen
  \bibfield  {author} {\bibinfo {author} {\bibfnamefont {M.}~\bibnamefont
  {Rieutord}}, \bibinfo {author} {\bibfnamefont {B.}~\bibnamefont {Georgeot}},
  \ and\ \bibinfo {author} {\bibfnamefont {L.}~\bibnamefont {Valdettaro}},\
  }\href@noop {} {\bibfield  {journal} {\bibinfo  {journal} {Phys. Rev. Lett.}\
  }\textbf {\bibinfo {volume} {85}},\ \bibinfo {pages} {4277} (\bibinfo {year}
  {2000})}\BibitemShut {NoStop}%
\bibitem [{\citenamefont {Maas}\ and\ \citenamefont {Lam}(1995)}]{ML95}%
  \BibitemOpen
  \bibfield  {author} {\bibinfo {author} {\bibfnamefont {L.}~\bibnamefont
  {Maas}}\ and\ \bibinfo {author} {\bibfnamefont {F.-P.}\ \bibnamefont {Lam}},\
  }\href@noop {} {\bibfield  {journal} {\bibinfo  {journal} {J. Fluid Mech.}\
  }\textbf {\bibinfo {volume} {300}},\ \bibinfo {pages} {1} (\bibinfo {year}
  {1995})}\BibitemShut {NoStop}%
\bibitem [{\citenamefont {Rieutord}\ \emph {et~al.}(2002)\citenamefont
  {Rieutord}, \citenamefont {Valdettaro},\ and\ \citenamefont
  {Georgeot}}]{RVG02}%
  \BibitemOpen
  \bibfield  {author} {\bibinfo {author} {\bibfnamefont {M.}~\bibnamefont
  {Rieutord}}, \bibinfo {author} {\bibfnamefont {L.}~\bibnamefont
  {Valdettaro}}, \ and\ \bibinfo {author} {\bibfnamefont {B.}~\bibnamefont
  {Georgeot}},\ }\href@noop {} {\bibfield  {journal} {\bibinfo  {journal} {J.
  Fluid Mech.}\ }\textbf {\bibinfo {volume} {463}},\ \bibinfo {pages} {345}
  (\bibinfo {year} {2002})}\BibitemShut {NoStop}%
\bibitem [{\citenamefont {{Vantieghem}}(2014)}]{vantieghem14}%
  \BibitemOpen
  \bibfield  {author} {\bibinfo {author} {\bibfnamefont {S.}~\bibnamefont
  {{Vantieghem}}},\ }\href@noop {} {\bibfield  {journal} {\bibinfo  {journal}
  {Proceedings of the Royal Society of London Series A}\ }\textbf {\bibinfo
  {volume} {470}},\ \bibinfo {pages} {40093} (\bibinfo {year}
  {2014})}\BibitemShut {NoStop}%
\bibitem [{\citenamefont {Poincar\'e}(1885)}]{Poinc1885}%
  \BibitemOpen
  \bibfield  {author} {\bibinfo {author} {\bibfnamefont {H.}~\bibnamefont
  {Poincar\'e}},\ }\href@noop {} {\bibfield  {journal} {\bibinfo  {journal}
  {Acta Mathematica}\ }\textbf {\bibinfo {volume} {7}},\ \bibinfo {pages} {259}
  (\bibinfo {year} {1885})}\BibitemShut {NoStop}%
\bibitem [{\citenamefont {Zhang}\ \emph {et~al.}(2001)\citenamefont {Zhang},
  \citenamefont {Earnshaw}, \citenamefont {Liao},\ and\ \citenamefont
  {Busse}}]{ZELB01}%
  \BibitemOpen
  \bibfield  {author} {\bibinfo {author} {\bibfnamefont {K.-K.}\ \bibnamefont
  {Zhang}}, \bibinfo {author} {\bibfnamefont {P.}~\bibnamefont {Earnshaw}},
  \bibinfo {author} {\bibfnamefont {X.}~\bibnamefont {Liao}}, \ and\ \bibinfo
  {author} {\bibfnamefont {F.}~\bibnamefont {Busse}},\ }\href@noop {}
  {\bibfield  {journal} {\bibinfo  {journal} {J. Fluid Mech.}\ }\textbf
  {\bibinfo {volume} {437}},\ \bibinfo {pages} {2001} (\bibinfo {year}
  {2001})}\BibitemShut {NoStop}%
\bibitem [{\citenamefont {Lebovitz}(1989)}]{lebovitz89}%
  \BibitemOpen
  \bibfield  {author} {\bibinfo {author} {\bibfnamefont {N.}~\bibnamefont
  {Lebovitz}},\ }\href@noop {} {\bibfield  {journal} {\bibinfo  {journal}
  {Geophys. Astrophys. Fluid Dyn.}\ }\textbf {\bibinfo {volume} {46}},\
  \bibinfo {pages} {221} (\bibinfo {year} {1989})}\BibitemShut {NoStop}%
\bibitem [{\citenamefont {{Cui}}\ \emph {et~al.}(2014)\citenamefont {{Cui}},
  \citenamefont {{Zhang}},\ and\ \citenamefont {{Liao}}}]{cui_etal14}%
  \BibitemOpen
  \bibfield  {author} {\bibinfo {author} {\bibfnamefont {Z.}~\bibnamefont
  {{Cui}}}, \bibinfo {author} {\bibfnamefont {K.}~\bibnamefont {{Zhang}}}, \
  and\ \bibinfo {author} {\bibfnamefont {X.}~\bibnamefont {{Liao}}},\
  }\href@noop {} {\bibfield  {journal} {\bibinfo  {journal} {Geophysical and
  Astrophysical Fluid Dynamics}\ }\textbf {\bibinfo {volume} {108}},\ \bibinfo
  {pages} {44} (\bibinfo {year} {2014})}\BibitemShut {NoStop}%
\bibitem [{\citenamefont {{Ivers}}\ \emph {et~al.}(2015)\citenamefont
  {{Ivers}}, \citenamefont {{Jackson}},\ and\ \citenamefont {{Winch}}}]{IJW15}%
  \BibitemOpen
  \bibfield  {author} {\bibinfo {author} {\bibfnamefont {D.~J.}\ \bibnamefont
  {{Ivers}}}, \bibinfo {author} {\bibfnamefont {A.}~\bibnamefont {{Jackson}}},
  \ and\ \bibinfo {author} {\bibfnamefont {D.}~\bibnamefont {{Winch}}},\
  }\href@noop {} {\bibfield  {journal} {\bibinfo  {journal} {J. Fluid Mech.}\
  }\textbf {\bibinfo {volume} {766}},\ \bibinfo {pages} {468} (\bibinfo {year}
  {2015})}\BibitemShut {NoStop}%
\bibitem [{\citenamefont {Rieutord}(2015)}]{rieutord15}%
  \BibitemOpen
  \bibfield  {author} {\bibinfo {author} {\bibfnamefont {M.}~\bibnamefont
  {Rieutord}},\ }\href@noop {} {\emph {\bibinfo {title} {{Fluid Dynamics: An
  Introduction}}}}\ (\bibinfo  {publisher} {Springer},\ \bibinfo {year}
  {2015})\ p.\ \bibinfo {pages} {508 pp.}\BibitemShut {Stop}%
\bibitem [{\citenamefont {Kellogg}(1953)}]{kellogg53}%
  \BibitemOpen
  \bibfield  {author} {\bibinfo {author} {\bibfnamefont {O.~D.}\ \bibnamefont
  {Kellogg}},\ }\href@noop {} {\emph {\bibinfo {title} {{Foundations of
  Potential Theory}}}}\ (\bibinfo  {publisher} {Dover, New York},\ \bibinfo
  {year} {1953})\BibitemShut {NoStop}%
\bibitem [{\citenamefont {Kudlick}(1966)}]{kudlick66}%
  \BibitemOpen
  \bibfield  {author} {\bibinfo {author} {\bibfnamefont {M.~D.}\ \bibnamefont
  {Kudlick}},\ }\emph {\bibinfo {title} {On transient motions in a contained
  rotating fluid}},\ \href@noop {} {Ph.D. thesis},\ \bibinfo  {school}
  {Massachusetts Inst. of Technology, Cambridge, Mass.} (\bibinfo {year}
  {1966})\BibitemShut {NoStop}%
\bibitem [{\citenamefont {Greenspan}(1965)}]{greenspan65}%
  \BibitemOpen
  \bibfield  {author} {\bibinfo {author} {\bibfnamefont {H.~P.}\ \bibnamefont
  {Greenspan}},\ }\href@noop {} {\bibfield  {journal} {\bibinfo  {journal} {J.
  Fluid Mech.}\ }\textbf {\bibinfo {volume} {22}},\ \bibinfo {pages} {449}
  (\bibinfo {year} {1965})}\BibitemShut {NoStop}%
\bibitem [{\citenamefont {Stewartson}\ and\ \citenamefont
  {Rickard}(1969)}]{SR69}%
  \BibitemOpen
  \bibfield  {author} {\bibinfo {author} {\bibfnamefont {K.}~\bibnamefont
  {Stewartson}}\ and\ \bibinfo {author} {\bibfnamefont {J.}~\bibnamefont
  {Rickard}},\ }\href@noop {} {\bibfield  {journal} {\bibinfo  {journal} {J.
  Fluid Mech.}\ }\textbf {\bibinfo {volume} {35}},\ \bibinfo {pages} {759}
  (\bibinfo {year} {1969})}\BibitemShut {NoStop}%
\bibitem [{\citenamefont {Greenspan}(1964)}]{greenspan64}%
  \BibitemOpen
  \bibfield  {author} {\bibinfo {author} {\bibfnamefont {H.~P.}\ \bibnamefont
  {Greenspan}},\ }\href@noop {} {\bibfield  {journal} {\bibinfo  {journal} {J.
  Fluid Mech.}\ }\textbf {\bibinfo {volume} {20}},\ \bibinfo {pages} {673}
  (\bibinfo {year} {1964})}\BibitemShut {NoStop}%
\bibitem [{\citenamefont {Lorch}(1962)}]{lorch62}%
  \BibitemOpen
  \bibfield  {author} {\bibinfo {author} {\bibfnamefont {E.~R.}\ \bibnamefont
  {Lorch}},\ }\href@noop {} {\emph {\bibinfo {title} {{Spectral Theory}}}}\
  (\bibinfo  {publisher} {Oxford, New York},\ \bibinfo {year}
  {1962})\BibitemShut {NoStop}%
\bibitem [{\citenamefont {{Amrouche}}\ \emph {et~al.}(1998)\citenamefont
  {{Amrouche}}, \citenamefont {{Bernardi}}, \citenamefont {{Dauge}},\ and\
  \citenamefont {{Girault}}}]{amrouche_etal98}%
  \BibitemOpen
  \bibfield  {author} {\bibinfo {author} {\bibfnamefont {C.}~\bibnamefont
  {{Amrouche}}}, \bibinfo {author} {\bibfnamefont {C.}~\bibnamefont
  {{Bernardi}}}, \bibinfo {author} {\bibfnamefont {M.}~\bibnamefont {{Dauge}}},
  \ and\ \bibinfo {author} {\bibfnamefont {V.}~\bibnamefont {{Girault}}},\
  }\href@noop {} {\bibfield  {journal} {\bibinfo  {journal} {Mathematical
  Methods in the Applied Sciences}\ }\textbf {\bibinfo {volume} {21}},\
  \bibinfo {pages} {823} (\bibinfo {year} {1998})}\BibitemShut {NoStop}%
\bibitem [{\citenamefont {Kreyszig}(1978)}]{kreyszig78}%
  \BibitemOpen
  \bibfield  {author} {\bibinfo {author} {\bibfnamefont {E.}~\bibnamefont
  {Kreyszig}},\ }\href@noop {} {\emph {\bibinfo {title} {{Introductory
  Functional Analysis with Applications}}}}\ (\bibinfo  {publisher} {{Wiley}},\
  \bibinfo {year} {1978})\BibitemShut {NoStop}%
\bibitem [{\citenamefont {Halmos}(1951)}]{halmos51}%
  \BibitemOpen
  \bibfield  {author} {\bibinfo {author} {\bibfnamefont {P.}~\bibnamefont
  {Halmos}},\ }\href@noop {} {\emph {\bibinfo {title} {{Introduction to Hilbert
  Spaces}}}}\ (\bibinfo  {publisher} {Chelsea, New York},\ \bibinfo {year}
  {1951})\BibitemShut {NoStop}%
\bibitem [{\citenamefont {Furuta}(2001)}]{furuta01}%
  \BibitemOpen
  \bibfield  {author} {\bibinfo {author} {\bibfnamefont {T.}~\bibnamefont
  {Furuta}},\ }\href@noop {} {\emph {\bibinfo {title} {{Invitation to linear
  operators}}}}\ (\bibinfo  {publisher} {{CRC Press}},\ \bibinfo {year}
  {2001})\BibitemShut {NoStop}%
\bibitem [{\citenamefont {Halmos}(1958)}]{halmos58}%
  \BibitemOpen
  \bibfield  {author} {\bibinfo {author} {\bibfnamefont {P.}~\bibnamefont
  {Halmos}},\ }\href@noop {} {\emph {\bibinfo {title} {{Finite-Dimensional
  Vector Spaces}}}}\ (\bibinfo  {publisher} {Van Nostrand, New York},\ \bibinfo
  {year} {1958})\BibitemShut {NoStop}%
\bibitem [{\citenamefont {Korevaar}(1968)}]{korevaar68}%
  \BibitemOpen
  \bibfield  {author} {\bibinfo {author} {\bibfnamefont {J.}~\bibnamefont
  {Korevaar}},\ }\href@noop {} {\emph {\bibinfo {title} {{Mathematical
  Methods}}}}\ (\bibinfo  {publisher} {Academic Press, New York},\ \bibinfo
  {year} {1968})\BibitemShut {NoStop}%
\bibitem [{\citenamefont {Courant}\ and\ \citenamefont {Hilbert}(1953)}]{CH53}%
  \BibitemOpen
  \bibfield  {author} {\bibinfo {author} {\bibfnamefont {R.}~\bibnamefont
  {Courant}}\ and\ \bibinfo {author} {\bibfnamefont {D.}~\bibnamefont
  {Hilbert}},\ }\href@noop {} {\emph {\bibinfo {title} {Methods of mathematical
  Physics}}}\ (\bibinfo  {publisher} {Interscience, New York},\ \bibinfo {year}
  {1953})\BibitemShut {NoStop}%
\bibitem [{\citenamefont {Cartan}(1922)}]{Cartan22}%
  \BibitemOpen
  \bibfield  {author} {\bibinfo {author} {\bibfnamefont {E.}~\bibnamefont
  {Cartan}},\ }\href@noop {} {\bibfield  {journal} {\bibinfo  {journal} {Bull.
  Sci. Math.}\ }\textbf {\bibinfo {volume} {46}},\ \bibinfo {pages} {317}
  (\bibinfo {year} {1922})}\BibitemShut {NoStop}%
\bibitem [{\citenamefont {Szeg\"o}(1967)}]{szego67}%
  \BibitemOpen
  \bibfield  {author} {\bibinfo {author} {\bibfnamefont {G.}~\bibnamefont
  {Szeg\"o}},\ }in\ \href@noop {} {\emph {\bibinfo {booktitle} {Colloquium
  Publications}}},\ \bibinfo {editor} {edited by\ \bibinfo {editor}
  {\bibfnamefont {R.~I.}\ \bibnamefont {Am. Math.~Soc.}, \bibfnamefont
  {Providence}}}\ (\bibinfo {year} {1967})\BibitemShut {NoStop}%
\bibitem [{\citenamefont {{Nurijanyan}}\ \emph {et~al.}(2013)\citenamefont
  {{Nurijanyan}}, \citenamefont {{Bokhove}},\ and\ \citenamefont
  {{Maas}}}]{nurijanyan_etal13}%
  \BibitemOpen
  \bibfield  {author} {\bibinfo {author} {\bibfnamefont {S.}~\bibnamefont
  {{Nurijanyan}}}, \bibinfo {author} {\bibfnamefont {O.}~\bibnamefont
  {{Bokhove}}}, \ and\ \bibinfo {author} {\bibfnamefont {L.~R.~M.}\
  \bibnamefont {{Maas}}},\ }\href@noop {} {\bibfield  {journal} {\bibinfo
  {journal} {Phys. Fluids}\ }\textbf {\bibinfo {volume} {25}},\ \bibinfo
  {pages} {126601} (\bibinfo {year} {2013})}\BibitemShut {NoStop}%
\bibitem [{\citenamefont {Kelvin}(1880)}]{kelvin1880}%
  \BibitemOpen
  \bibfield  {author} {\bibinfo {author} {\bibfnamefont {L.}~\bibnamefont
  {Kelvin}},\ }\href@noop {} {\bibfield  {journal} {\bibinfo  {journal} {Phil.
  Mag.}\ }\textbf {\bibinfo {volume} {10}},\ \bibinfo {pages} {155} (\bibinfo
  {year} {1880})}\BibitemShut {NoStop}%
\end{thebibliography}%

\appendix

\section{$\pm1$ cannot be eigenvalues of the Poincar\'e problem}
Let us consider the momentum equation and its complex conjugate, namely
from \eq{inert}

\[ -\lambda\vv+ i\hat{\bf\Omega}\times\vv =-i\nabla q \andet
-\lambda\vv^*- i\hat{\bf\Omega}\times\vv^* = i\nabla q^*\]
where $\lambda=\pm1$. Let multiply the equations together. Hence, we get

\beq \|\nabla q\|^2 = -\|\vv\|^2 + \|\hat{\bf\Omega}\times\vv\|^2 +
i\lambda(\vv^*\cdot\na q - \vv\cdot\na q^*)\eeq
where we used $\lambda^2=1$ and the equations a second time. Noting that

\begin{equation} \|\hat{\bf\Omega}\times\vv\|^2 = \|\vv\|^2 -
\left|\dz{q}\right|^2
\end{equation}
where we aligned the rotation axis with the $z$-axis. Thus,
we obtain

\beq \|\nabla q\|^2+\left|\dz{q}\right|^2 = i\lambda(\vv^*\cdot\na q -
\vv\cdot\na q^*)\eeq
which we now integrate over the fluid volume. We finally obtain

\beq \intvol \|\nabla q\|^2+\left|\dz{q}\right|^2 dV = 0 \eeqn{nogradp}
where we used that

\[ \intvol \vv^*\cdot\na q\; dV= 0\]
which trivially follows from mass conservation and boundary conditions
when the velocity field is differentiable, but which is also true for
merely square-integrable velocity fields thanks to \eq{eq36} since
$\vv^*\in\ubLam$.

Hence, from \eq{nogradp}, we find that $\nabla q=\vzero$. Now we need to
check that the vanishing pressure gradient implies a vanishing velocity
field. From the equations of motion, we immediately find that

\beq v_z = 0 \andet v_y=\pm iv_x \eeqn{vxvy}
So the motion, if it
exists, is only a planar flow, perpendicular to the rotation axis.

Then, mass conservation demands that $\vv\in\ubLam$ (cf Eq. \ref{eq38}), which
means that for every $\phi\in\Pi^\infty$ we have

\beq\intvol \vv\cdot\na\phi^*\; dV = 0 \eeq
With \eq{vxvy}, setting $f=\partial_x\phi-i\partial_y\phi$, it also
means that for any $f\in\Pi^\infty$ we have

\beq \intvol v_x f^*\; dV = 0 \eeq
Thus $v_x$ is orthogonal to all infinitely differentiable complex scalar
functions defined on the volume $V$. It can only be zero, and so is the
velocity field. Hence,

\bigskip
\centerline{\fbox{$\pm1$ are not eigenvalues of the Poincar\'e
problem.}}

\bigskip
Let us now comment this mathematical result from a more physical view
point. The fact that the numbers $\pm1$ are excluded from the eigenvalue
spectrum comes from the fact that the fluid's domain is bounded. To view
that, it suffices to consider the propagation of characteristics that
are associated with the Poincar\'e operator. In a meridional section of
the fluid's volume, these characteristics are straight lines that bounce
on the boundaries \cite[e.g. fig. 8 or 9 in][]{RV97}. When the frequency
gets close to unity, the characteristics get almost perpendicular to the
rotation axis and, as they bounce on the boundaries, they form a web of
lines which is very dense. If we recall that characteristic lines are
the trace of equiphase surfaces, we understand that phase oscillates
very rapidly in the z-direction. In other words the wavenumber $k_z$
tends to infinity. Thus no mode can exist at $\lambda=\pm1$ while there
is no impediment for a propagating wave in the direction parallel to
the rotation axis in an unbounded domain.

\section{Explicit form of the Poincar\'e modes in the axisymmetric
ellipsoid}

Suppose that $E$ is an ellipsoid symmetric about the axis of rotation of
the fluid.  Choose Cartesian coordinates $x, y, z$ with $z$ along the axis
of rotation. Thus the unit vector $ {\hat {\bf z}} $ in the $z$ direction
is $ {\hat{\bf\Omega}} $ and the boundary $ \partial E $ has the equation

\begin{equation}
{x^2 \ +\  y^2   \over a^2 } \ +\  {z^2   \over c^2 } \ =\  1
\label{eq61}\end{equation}
where $a$ and $c$ are the two semiaxes of $ \partial E $.  It is
convenient to introduce cylindrical polar coordinates $s$, $\phi$, $z$
where $s\ =\ (x^2 \ +\ y^2 )^{1/2} $ and $ x\ =\ s\, \cos \, \phi \,,\ \
y\ =\ s\, \sin \, \phi $.  In these coordinates, the longitude $ \phi $
separates and we may seek solutions of \eq{poincpb} in the form

\begin{equation}
q\ =\ e^{{im} \phi } \  \tilde q (s,\,z)
\label{eq62}\end{equation}
where $m$ is any integer.  Substituting \eq{eq62} in \eq{poincpb} gives

\begin{subequations}
\begin{equation}
\begin{array}{l}
(s \partial_s \ +\  m\, \lambda^{-1}) \tilde q  \\
\\ \qquad\qquad =  (a/c)^2
( \lambda^{-2} -1)z \partial_z \tilde q\, \qquad {\rm on}\quad  \partial E
\end{array}
\eeqn{eq63a}
\begin{equation}
\begin{array}{l}
(\partial_s^2 \ +\  s^{-1} \partial_s - m^2 s^{-2}) \tilde q  \\
\\ \qquad\qquad=(\lambda^{-2} -1)\partial_z^2 \tilde q \qquad {\rm in} \quad  E.
\end{array}
\eeqn{eq63b}
\label{eq63}
\end{subequations}

For any $ \lambda $ satisfying 
\begin{equation}
0 < |\lambda| < 1
\label{eq64}\end{equation}
Bryan (1889) sought a solution of \eq{eq63} by introducing a 
system of confocal spheroidal coordinates depending on and 
adapted to that particular value of $ \lambda $.  Bryan's 
coordinate systems are most simply described in trigonometric 
terms.  Given a $ \lambda $ satisfying \eq{eq64}, choose 
$ \gamma $ so that 

\begin{subequations}
\begin{equation}
0 <  | \gamma | < \pi /2 
\label{eq65a}\end{equation}
\begin{equation}
\tan\gamma \ =\ (c/a)\, \lambda \,(1\ -\  \lambda^2 )^{-1/2} \,.
\label{eq65b}\end{equation}
\label{eq65}
\end{subequations}
(In this paper, when $x > 0$ then $x^{1/2} $ is always the
positive square root of $x$.)  Given $ \gamma $, we can recover
$ \lambda $ as 

\begin{subequations}
\begin{equation}
\lambda \ =\ \frac{\sin\gamma}{h(\gamma)}
\label{eq66a}\end{equation}
where
\begin{equation}
h( \gamma )\ =\ a^{-1} \,(\,a^2 \, \sin^2 \gamma \ +\  c^2
\,\cos^2 \gamma \,)^{1/2} \,.
\label{eq66b}\end{equation}
\label{eq66}
\end{subequations}

To obtain the trigonometric version of Bryan's curvilinear
coordinates, in the ($s, z$) plane consider the half-ellipse
\begin{equation}
s^2 / a^2 \ +\  z^2 / c^2 \ =\ 1\,;\ \ \ \ \ \ s\ge0
\label{eq67}\end{equation}
obtained from \eq{eq61}.  For any $ \gamma $ satisfying \eq{eq65a},
let $( \xi ,\, \eta )$ be curvilinear coordinates inside \eq{eq67},
chosen so that 

\begin{subequations}
\begin{equation}
s\ =\ a\  { \cos \xi \cos \eta \over  \cos \gamma }
\label{eq68a}\end{equation}
\begin{equation}
z\ =\ c\  { \sin \xi  \sin \eta   \over  \sin \gamma }
\label{eq68b}\end{equation}
with 
\begin{equation}
| \gamma | <  \xi < \pi /2
\label{eq68c}\end{equation}
and 
\begin{equation}
-|\gamma| < \eta < |\gamma |\,.
\label{eq68d}\end{equation}
\label{eq68}
\end{subequations}

\begin{figure}
\begin{center}
 \includegraphics[width=0.8\linewidth]{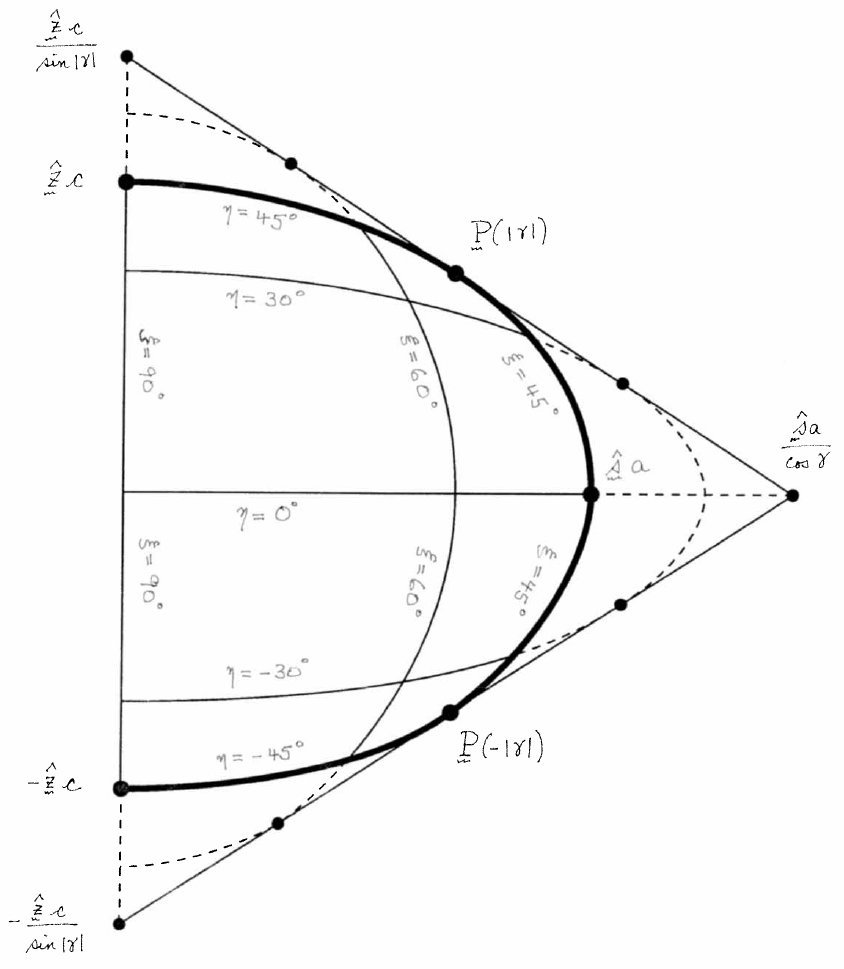}
 \caption{Bryan's ellipsoidal coordinate system \eq{eq68} when
$\gamma=45^\circ$ or $\gamma=-45^\circ$ and $2a=3c$. The fluid lies
inside the heavy ellipse where either $\eta=-|\gamma|$, or
$\xi=|\gamma|$, or $\eta=|\gamma|$. The points ${\bf P}(\pm|\gamma|)$ are
given by equation \eq{eq69}, $\hat{\bf s}$ and $\hat{\bf z}$ being unit
vectors in the direction of increasing $s$ and $z$.}
   \label{fig1}
\end{center}
\end{figure}

Figure 1 shows the curvilinear coordinate system $( \xi ,\, \eta )$
generated by a typical $ \gamma $ satisfying \eq{eq65a}.  In that
figure, the two oblique straight lines are drawn so as to be
tangent to \eq{eq67} at the points $ {\bf P} ( \pm \gamma )$,
where
\begin{equation}
{\bf P} ( \gamma )\ =\ {\hat {\bf s}} \,a\, \cos \, \gamma \ +\ 
{\hat {\bf z}} \,c\, \sin \, \gamma \,,
\label{eq69}\end{equation}
${\hat {\bf s}} $ being the unit vector in the $s$ direction in the $(s,
z)$ plane.  All the level curves of $ \xi $ and $ \eta $ obtained from
\eq{eq68} are arcs of half-ellipses tangent to those two oblique lines.
The level curves $ \xi $ = constant belong to half-ellipses which
intersect \eq{eq67} between $ {\bf P} ( \pm \gamma )$ and the $z$ axis,
while the level curves $ \eta $ = constant belong to the half-ellipses
which intersect \eq{eq67} between ${\bf P} ( \pm \gamma )$ and $a {\hat
{\bf s}} $.  The level curve $ \xi \ =\  \pi /2$ is the segment of
the $z$-axis connecting $-c {\hat {\bf z}} $ and $c {\hat {\bf z}} $.
The level curve $ \xi \ =\ | \gamma |$ is the part of \eq{eq67} connecting $
{\bf P} ( \gamma )$ and $ {\bf P} (- \gamma )$.  The level curve $ \eta \
=\ -| \gamma |$ is the part of \eq{eq67} connecting $-c\, {\hat {\bf z}} $
and $ {\bf P} (-| \gamma |)$.  The level curve $ \eta \ =\ | \gamma |$
is the part of \eq{eq67} connecting $c\, {\hat {\bf z}} $ and $ {\bf P} (|
\gamma |)$.  The level curve $ \eta \ =\ 0$ is the segment of the $s$
axis connecting the origin and $a{\bf\hat s} $.

In terms of the coordinates $( \xi , \eta )$ the partial derivatives 
$ \partial_s $ and $ \partial_z $ are as follows:
\begin{subequations}
\begin{equation}
aD(\xi, \eta) \sec\gamma \partial_s = \sin\xi\cos\eta
\partial_{\xi} - \cos\xi\sin\eta\partial_{\eta}
\label{eq610a}\end{equation}
\begin{equation}
cD( \xi , \eta ) \csc  \gamma \partial_z = \cos\xi\sin \eta
\partial_{\xi} - \sin \xi \cos\eta \partial_{\eta}
\label{eq610b}\end{equation}
where
\begin{equation}
D(\xi , \eta )\ =\  \cos^2 \xi - \cos^2 \eta \,.
\label{eq610c}\end{equation}
\label{eq610}
\end{subequations}

Straightforward calculation using \eq{eq65b} then shows that

\begin{subequations}
\beqan
(a \sec\gamma )^2 \,D( \xi , \eta )\,[ \partial_s^2 \ +\ s^{-1}
\, \partial_s \ -\nonumber \\  m^2 s - (\lambda^{-2}-1)\partial_z^2 ] \ =\
L_{\eta}^{(m)} \ -\  L_{\xi}^{(m)} \eeqan{eq611a}
where

\begin{equation}
L_{\eta}^{(m)} \ =\  \partial_{\eta}^2 \ -\  \tan\eta \,
\partial_{\eta} \ -\  m^2 ( \sec\eta )^2 \,.
\label{eq611b}\end{equation}
\label{eq611}
\end{subequations}
In the same way, 

\begin{equation}
\begin{array}{l}
D(\xi,\eta)\sin^2\gamma\left[s\partial_s + m
\lambda^{-1} - (a/c)^2 (\lambda^{-2} - 1)z \partial_{z}\right] \\ \\
= D(\eta,\gamma)\sin\xi\cos\xi\partial_\xi  +
D(\gamma,\xi)\sin\eta\cos\eta\partial_\eta  \\ \\
- D(\eta,\xi)m \lambda^{-1}\sin^2\gamma .
\end{array}
\label{eq612}\end{equation}

\par\noindent
Thus the Poincar\'e equation \eq{eq63b} becomes
\begin{equation}
L_{\eta}^{(m)} \, \tilde q \ =\  L_{\xi}^{(m)} \, \tilde q
\label{eq613}\end{equation}
\par\noindent
and the boundary condition \eq{eq63a} separates into three parts
corresponding to the three arcs into which $ {\bf P} ( \gamma )$
and $ {\bf P} (- \gamma )$ divide the half-ellipse \eq{eq67}.  To
satisfy \eq{eq63a} $ \tilde q $ must behave as
follows: for $|\gamma| < \xi < \pi/2$ one must have

\begin{subequations}
\begin{equation}
\begin{array}{l}
[\sin\eta\cos\eta\partial_\eta - mh(\gamma)\sin\gamma]\tilde q (\xi,\eta) = 0
\\ \\ {\rm at} \ \ \  \eta \ =\  \pm \gamma \,;
\end{array}
\label{eq614a}\end{equation}
and for $-|\gamma| < \eta < |\gamma|$ one must have
\begin{equation}
\begin{array}{l}
[\sin\xi\cos\xi \partial_\xi - mh(\gamma) \sin\gamma]\tilde q (\xi, \eta)=0
\\ \\
 {\rm at} \ \ \  \xi \ =\  | \gamma |\,.
\end{array}
\label{eq614b}\end{equation}
\label{eq614}
\end{subequations}

Because of \eq{eq613}, a particular solution of \eq{eq63b} can be obtained
by choosing any integer $l\, \ge \,|\,m|$ and setting 
\begin{subequations}
\begin{equation}
\tilde q ( \xi , \eta )\ =\ P_l^m ( \sin \, \xi )\,
P_l^m ( \sin \, \eta )
\label{eq615a}\end{equation}
where $P_l^m $ is the associated Legendre function,
\begin{equation}
P_l^m ( \mu )\ =\ (2^l \, l!)^{-1} \,(1- \mu^2 )^{m/2} 
\, \partial_{\mu}^{l+m} \,( \mu^2 \ -\ 1)^l \,.
\label{eq615b}\end{equation}
\label{eq615}
\end{subequations}
\par\noindent
This $\tilde q $ will also satisfy the boundary conditions \eq{eq63a} if
it satisfies \eq{eq614}, that is, if
\begin{subequations}
\begin{equation}
\begin{array}{l}
[\sin\eta\cos\eta\partial_\eta - mh(\gamma)\sin\gamma] 
P_l^m (\sin\eta)  = 0 \\
\\
{\rm at} \ \ \  \eta \ =\  \pm \gamma
\end{array}
\label{eq616a}\end{equation}
and also
\begin{equation}
\begin{array}{l}
[\sin\xi\cos\xi \partial_{\xi}  -  mh(\gamma)\sin\gamma] 
P_l^m (\sin\xi)  = 0 \\ \\ 
{\rm at} \ \ \  \xi \ =\ | \gamma |\,.
\end{array}
\label{eq616b}\end{equation}
\label{eq616}
\end{subequations}
Obviously \eq{eq616a} implies \eq{eq616b}.  Moreover, the left side of
\eq{eq616a} has the same parity in $ \eta $ as does $P_l^m ( \sin 
\, \eta )$, so if \eq{eq616a} is satisfied for $ \eta \ =\  \gamma $ it is
also satisfied for $ \eta \ =\ - \gamma $.  At $ \eta \ =\  \gamma $,
\eq{eq616a} reduces to
\begin{subequations}
\begin{equation}
[\, \cos \, \eta \, \partial_{\eta} \ -\ mh( \eta )\,]\ P_l^m ( \sin \, \eta )
\ =\ 0
\label{eq617a}\end{equation}
where, because of \eq{eq65a},
\begin{equation}
0 < | \eta | <  \pi/2\,.
\label{eq617b}\end{equation}
\label{eq617}
\end{subequations}
Note that when $m\ =\ 0$ the choice $l\ =\ 0$ is of no interest because
then in \eq{eq615a} $\tilde q \ =\ 1$ so \eq{vvp} gives ${\bf v} = {\bf 0} $.

Now we can summarize Bryan's (1889) recipe for constructing some
eigenfunctions $q$ and their corresponding eigenvalues $ \lambda $ in
the Poincar\'e pressure problem \eq{poincpb}:   choose any integer $l\  \ge \ 1$
and any integer $m$ satisfying $-l\  \le \ m\  \le \ l$.  Find a root $
\eta $ of \eq{eq617} and set $ \gamma = \eta $.  Then use this $ \gamma $
to generate a curvilinear coordinate system \eq{eq68} inside the fluid
ellipsoid.  Choose

\begin{equation}
q(s,\, \phi \,,\,z)\ =\ e^{{im} \phi } \,P_l^m ( \sin \, \xi )
\,P_l^m ( \sin \, \eta )
\label{eq618}\end{equation}
where $s$ and $z$ are given by \eq{eq68}.  Finally, calculate 
$ \lambda $ from $ \gamma $ via \eq{eq66}.

\end{document}